\documentclass[sigconf, nonacm]{acmart}

\AtBeginDocument{%
  }

\begin{document}

\title{Automating Autograding: Large Language Models as Test Suite Generators for Introductory Programming}

\author{Umar Alkafaween}
\orcid{0009-0004-6003-3370}
\affiliation{%
  \institution{Al Hussein Technical University}
  \city{Amman}
  \country{Jordan}
}
\email{umar.alkafaween@htu.edu.jo}

\author{Ibrahim Albluwi}
\orcid{0000-0003-1816-3943}
\affiliation{%
  \institution{Princess Sumaya University for Technology}
  \city{Amman}
  \country{Jordan}
}
\email{i.albluwi@psut.edu.jo}

\author{Paul Denny}
\orcid{0000-0002-5150-9806}
\affiliation{%
  \institution{University of Auckland}
  \state{Auckland}
  \country{New Zealand}
}
\email{paul@cs.auckland.ac.nz}

\renewcommand{\shortauthors}{Alkafaween et al.}

\begin{abstract}
Automatically graded programming assignments provide instant feedback to students and significantly reduce manual grading time for instructors. However, creating comprehensive suites of test cases for programming problems within automatic graders can be time-consuming and complex. The effort needed to define test suites may deter some instructors from creating additional problems or lead to inadequate test coverage, potentially resulting in misleading feedback on student solutions. Such limitations may reduce student access to the well-documented benefits of timely feedback when learning programming.

In this work, we evaluate the effectiveness of using Large Language Models (LLMs), as part of a larger workflow, to automatically generate test suites for CS1-level programming problems.  
Each problem’s statement and reference solution are provided to GPT-4 to produce a test suite that can be used by an autograder.  We evaluate our proposed approach using a sample of 26 problems, and more than 25,000 attempted solutions to those problems, submitted by students in an introductory programming course. 
We compare the performance of the LLM-generated test suites against the instructor-created test suites for each problem.
Our findings reveal that LLM-generated test suites can correctly identify most valid solutions, and for most problems are at least as comprehensive as the instructor test suites.
Additionally, the LLM-generated test suites exposed ambiguities in some problem statements, underscoring their potential to improve both autograding and instructional design.
\end{abstract}





\maketitle

\section{Introduction}\label{sec1}

Autograding of programming assignments offers a number of well-documented benefits to both students and instructors. In particular, the instant feedback provided by autograders has been shown to help students correct their errors and solve more problems \cite{pieterse2013automated,sherman2013impact,duch2018dante}. Students perceive instant feedback as a positive contribution to their learning \cite{farnqvist2016competition,rao2019experiences}, and this is reflected in their improved overall performance \cite{gordillo2019effect, bai2016enhancing}. Autograding also minimizes the variability in grading decisions which may be subjective when grading is performed by instructors or teaching assistants \cite{parihar2017automatic}, and it helps reduce the workload of teaching staff significantly, by eliminating the need to manually evaluate each submission \cite{bai2016enhancing, venables2003programming}. 
However, preparing problems for autograding systems requires more than just a problem statement; instructors must also provide a test suite that thoroughly covers the different scenarios of the problem. Preparing this test suite is not always straightforward. For it to be thorough, it must at least contain tests for multiple random inputs, the minimum and maximum limits of the problem's inputs, as well as other edge-cases that are dependent on the problem. Furthermore, some problems require unique tests that capture specific constraints in the problem, e.g., an array of distinct numbers, or an array of non-increasing numbers, and sometimes random tests must be validated to ensure they actually contain a valid answer for the problem. This can all be very time consuming, even for low-level courses \cite{lobb2016coderunner}. Moreover, failing to provide sufficient tests might result in misleading feedback to students, as invalid solutions could be graded as valid.

Recent advances in Generative AI (GenAI) have introduced exciting new possibilities for using AI in Computing Education \cite{denny2024computing}. Large Language Models (LLM) in particular, which are capable  of generating human-like outputs in response to text prompts, can be integrated in various Computing Education settings \cite{becker2023programming, 10.1145/3623762.3633499}. They have been applied to generating new programming exercises \cite{sarsa2022automatic}, explaining code \cite{denny2024desirable, liffiton2023codehelp, leinonen2023comparing, phung2023generative}, providing feedback and improving the readability of programming error messages \cite{leinonen2023using, phung2023generative}, and powering novel pedagogical approaches \cite{smith2024prompting, denny2024prompt}.  While they have also been used to grade solutions based on a set of criteria provided to the LLM \cite{nilsson2023gpt, bengtsson2023assessment}, we are not aware of prior work that uses LLMs in a Computing Education context to help instructors generate test suites for autograders to address the challenges stated earlier.

In this study, we explore the efficacy of using LLMs to minimize the time and effort required of instructors who want their assignments to be submitted to and evaluated by autograders. We run tens of thousands of student solutions on LLM-generated test suites for CS1 problems, and compare their results to running them on traditional instructor-generated test suites with respect to their correctness and thoroughness. We also look into the LLM-generated test suites' ability to uncover ambiguities in instructor-written problem statements. Our analysis is guided by the following three research questions:

\begin{enumerate}[]
\item \textbf{RQ1:} To what extent do LLM-generated test suites correctly identify valid solutions to CS1 problems?
\item \textbf{RQ2:} How comprehensive are LLM-generated test suites compared to instructor-generated test suites?
\item \textbf{RQ3:} What types of ambiguities can LLM-generated test suites help uncover in problem statements?
\end{enumerate}

We start with a literature review of studies on autograding and LLMs in Computing Education. We also show how our work differs from the large field of work undertaken on generating unit-tests for code. Then we describe the design of our study, present the results, and discuss the broader implications of this work.

\section{Background}\label{sec2}

Researchers agree that introductory programming courses should not only focus on teaching syntax, but also help students build mental scaffolds that allow them to follow problem solving processes and develop awareness of where they are in such a process \cite{prather2018metacognitive}. The ability to use these scaffolds and exercise cognitive control to progress through the problem solving process is defined as self-regulation and is critical to the learning process \cite{prather2020we}. Providing feedback to students is known to help with self-regulation \cite{prather2018metacognitive,hao2019automated}, but the growing number of students enrolling in programming courses makes the process of grading each student solution manually inefficient. To adapt to this growth, instructors frequently rely on tools to automate the grading of assignments \cite{hao2019automated,keuning2018systematic}.

There are a large number of tools available for autograding programming assignments \cite{keuning2018systematic,paiva2022automated}, and autograding can be carried out at different levels. Static analysis methods depend only on the source code and do not require its execution. They can detect issues in code functionality, quality, safety, syntax, and some runtime errors \cite{staubitz2015towards,skalka2023development}. In contrast, dynamic analysis methods require the execution of the source code and are the most effective way to assess its functionality \cite{paiva2022automated}. Most of these autograding tools automate the process of submission and grading, but still require instructors to prepare tests for the assignment problems beforehand \cite{keuning2018systematic}. The tools use these tests to produce a grade for each submission. This grade can be binary, i.e., pass/fail, or partial as a percentage of the number of tests the submission passed.

One static approach that automates generating tests and grading student solutions is introduced by LEGenT \cite{agarwal2022legent}. LEGenT works offline by analysing solutions and clustering them into good and bad solutions according to their similarity to a reference solution. Solutions in the clusters are further clustered according to similarity in their structure and semantics. A representative solution from a bad cluster is picked and a good solution with similar structure is picked from a good cluster. A test is generated through a number of steps carried out on the good and bad representatives. This test can be used to test all the solutions in the bad cluster. A more complex approach is used for running LEGenT in an online scenario to test solutions as they are submitted. LEGenT does not support advanced constructs like arrays, pointers, and recursion, and requires a few manual tests to be created by the instructor for clustering.

A mix of dynamic analysis methods was used by Skalka and Drlík (2023) \cite{skalka2023development} to develop tests to evaluate student source code in object oriented programming problems. They design classes that can be used to generate random tests based on simple configurations, and compare the outputs of the student solutions with that of a reference solution. While their approach provides a simpler and structured way to create tests, it still requires some effort from the instructor to adapt the test cases to each different programming task.as

In a recent literature review, Messer et al. (2023) \cite{messer2023machine} reviewed 27 papers that used Machine Learning techniques to grade and provide feedback on programming assignments. They define four fundamental programming attributes: correctness, maintainability, readability and documentation. They report that 56\% of the papers focus on assessing correctness, and 61\% of the tools presented in the papers use a neural-network approach. On average, the tools reported an accuracy of at least 70\% in predicting the correct grades or giving correct feedback.

Research on Artificial Intelligence in Computing Education surged since November 2022 with the introduction of ChatGPT, which drew widespread attention from the public and researchers to Large Language Models (LLMs). Prather et al. (2023) \cite{10.1145/3623762.3633499} highlight a few ways LLMs can be used to anaylse student work, such as bug fixing, enhancing programming error messages, grading assessments, and providing feedback. We highlight a few studies that explored the effectiveness of LLMs in grading student submissions and providing feedback on programming assignments.

Phung et al. (2023) \cite{phung2023generative} compared GPT-3.5 and GPT-4 to human tutors in multiple scenarios like program repair, hint generation, and grading. They supply the problem statement, grading rubric, and student solution in Python to the LLM, and instruct it to generate a grade for the solution based on the rubric. Their results show that GPT-4 can come close to human tutors in some scenarios, but performs significantly worse in grading compared to human tutors. Their assumption is that testing edge-cases requires in-depth reasoning that the LLM may not be good at.

To test LLMs' capabilities in grading compared to teaching assistants, Nilsson and Tuvstedt (2023) \cite{nilsson2023gpt} used data from an introductory programming course and supplied GPT-4 with the assignment instructions, grading criteria, grading instructions, and student submissions made in Java and other programming languages for 73 students. They concluded that the LLM exhibited 75\% grading accuracy when compared to grades given previously on the same submissions by teaching assistants, but it performed significantly worse at identifying failing submissions compared to passing ones.

Similarly, Bengtsson and Kaliff (2023) \cite{bengtsson2023assessment} used GPT-4 to inject four types of logical errors into correct Java student submissions in four Data-structures tasks, and evaluated the grades the LLM produced for the original and modified solutions. Their results showed that GPT-4 was able to detect correct submissions 90\% of the time on average when the LLM is provided with task instructions. Not providing task instructions reduced the accuracy of detecting correct submissions to 66\% on average, but improved the accuracy of detecting submissions with errors.

Azaiz et al. (2024) \cite{10.1145/3649217.3653594} evaluated the feedback GPT-4 Turbo produced on 55 student submissions in Java. Given the task specification and student submission, the LLM was prompted to find all the errors in the submission and provide hints for correcting them. They report that the feedback generated by GPT-4 Turbo is personalized for each submission, mainly because the feedback does not rely on test cases, and that 52\% of the submissions received fully correct and complete feedback compared to only 31\% by GPT-3.5 in previous studies on the same dataset \cite{azaiz2023ai}. While GPT-4 Turbo provided better feedback than its predecessor, it still generated incomplete feedback or feedback that contained redundancies, inconsistencies, or unclear explanations. They concluded that using GPT-4 Turbo to provide automatic feedback for student submissions is not advisable as it may increase the cognitive load on students, but using it as a tool to aid teaching assistants in grading could prove efficient.

The aforementioned studies focus on evaluating the capabilities of LLMs in grading solutions. We are not aware of prior studies that focus on using LLMs to generate tests to be used in autograding systems.
One large area of research that may appear similar to the work in this study is that of generating unit tests in the Software Engineering field. There is a growing body of research on using LLMs to generate unit tests in this area \cite{siddiq2024using, wang2024software, schafer2023empirical, chen2023teaching}, our work, however, differs from it in that:

\begin{enumerate}[1.]
\item It bases the generation of unit tests on a reference solution only, but we also use the problem statement. This allows us to address requirements that may not be clear in the reference solution. For example: requiring the presence or absence of certain functions or features.

\item Evaluation metrics in Software Engineering do not necessarily align with those in Computing Education. In Computing Education, the priority is to have a functional set of tests suitable for autograding purposes. In Software Engineering, however, the focus is more on meeting standards for coverage, maintainability, extensibility, and other quality attributes.
\end{enumerate}

\section{Methodology}\label{sec3}

\subsection{Data}

\subsubsection{Problems}

We used 28 problems taken from seven labs in an introductory course on Engineering Computation and Software Development taught in Fall 2023 at The University of Auckland. All problems are solved using the C programming language. 8 problems required students to write full C programs with a main function that accepts input from Standard Input and prints output to Standard Output. These problems are referred to in the study as full-program problems. 18 problems required students to implement a function that returns data and/or prints output to Standard Output. These problems are referred to in the study as function-implementation problems. The two remaining problems either accepted input from files or used files to find the solution. These two problems were excluded from the study because handling files was out of scope. Appendix \ref{app1} shows samples of the problems. All 26 problems (8 full-program, 18 function-implementation) had the following relevant parts:
\begin{enumerate}[]
\item \textbf{Problem statement:} contains the description of the problem and sometimes contains images.
\item \textbf{Reference solution:} a valid solution for the problem, provided by the instructor in C.
\item \textbf{Extra code (optional):} contains any starter code provided to students with the problem statement or made available to their code to use. This may include C preprocessing directives, global variables, and helper function definitions.
\item \textbf{A list of instructor tests:} For full-program problems, a test is specified as an input that will be fed to the student solution from Standard Input. For function-implementation problems, the test is specified as C code that calls the function implemented by the student solution. All tests also specify an expected output that is compared to the output actually produced by the student solution to verify whether or not it is valid.
\end{enumerate}

\begin{table*}[!h]%
    \caption{Summary of used problems and number of solutions for each problem. Problems 14 and 21 (in red) were excluded from the study because they use files.
    \label{tab:table1}}
    \begin{tabular*}{\textwidth}{p{1cm} p{12.74cm} p{3cm}}\toprule
        \textbf{Problem \#} & \textbf{Problem description summary} & \textbf{\# of solutions} \\ \midrule
        1 & Function to find the number of ways to choose m things out of a total of n things (combinations) & 2,188 \\ \hline
        2 & Function to update a 2D array to show movement from a source (marked with value 1) to a destination (marked with value 2). Movement should be done horizontally then vertically & 986 \\ \hline
        3 & Function to return a number times two & 1,168 \\ \hline
        4 & Print a couple of lines with static text & 851 \\ \hline
        5 & Accept a letter and print the letter after it & 717 \\ \hline
        6 & Function to reverse a number's digits recursively & 2,067 \\ \hline
        7 & Function to check if the hamming distance between two strings is one & 1,384 \\ \hline
        8 & Function to print an array & 1,343 \\ \hline
        9 & Accept a letter and a number of steps then print the letter that comes after moving that amount of steps from the input letter & 1,641 \\ \hline
        10 & Function to find if a given string is a palindrome & 1,460 \\ \hline
        11 & Print ``number $\times$ 2 = result'' replace number and result with the actual values & 1,532 \\ \hline
        12 & Print the cross-sectional area of a pipe & 948 \\ \hline
        13 & Function to print a number times two & 1,475 \\ \hline
        \rowcolor{pink}
        14 & Read numbers from a file, group them into buckets and print a histogram for the buckets using the letter X for each value in a bucket & 2,836 \\ \hline
        \rowcolor{white}
        15 & Function to count zeros in an array & 1,210 \\ \hline
        16 & Function to substitute letters in a string from a starting index to an ending index. Both indices are always valid and the starting index is always less than or equal to the ending index & 1,170 \\ \hline
        17 & Function to capitalize words in a sentence & 1,740 \\ \hline
        18 & Hello world & 951 \\ \hline
        19 & Function to find a number in a 2D array and set its row and column indices to addresses that are passed to the function pointer parameters & 1,158 \\ \hline
        20 & Function to find the last occurrence of a number in an array & 1,231 \\ \hline
        \rowcolor{pink}
        21 & Function that increases the sum of digits of two numbers each to itself. Stops after X iterations of doing this and returns the max between the two numbers, or stops if in any iteration, a number becomes equal to any value the other number was equal to in any iteration, returns that value & 755 \\ \hline
        \rowcolor{white}
        22 & Function to find the two in-order numbers with the largest difference and the shortest distance between their indices in an array & 916 \\ \hline
        23 & Function to convert an integer into its binary string recursively & 2,096 \\ \hline
        24 & Find the total number of containers from two different sizes needed to fulfill a shipping order of some size & 871 \\ \hline
        25 & Function to reverse an array in place & 1,385 \\ \hline
        26 & Function to draw a histogram for elements in an array using ``ASCII art''. Function should concatenate the characters of the drawing to a string that is passed to the function & 705 \\ \hline
        27 & Function to find the midpoint between two points & 993 \\ \hline
        28 & Function to swap two adjacent elements in a 2D array if possible & 1,563 \\ 
        \bottomrule
    \end{tabular*}
\end{table*}

\subsubsection{Student Solutions}

We used a total of 33,749 solutions submitted by students during the semester for all 26 problems.  The course uses the CodeRunner autograder \cite{lobb2016coderunner}, and for each problem the submissions were exported as a CSV file using a custom script. The format of this file lists each attempt as a row that includes the submitted code for that attempt, a universal anonymized ID for the student that made the attempt, the time and date of the attempt, and a binary digit that is either 1 if the attempt solved the problem correctly, or 0 otherwise. In this exported data, an attempt is considered to have solved the problem correctly if it produced the expected output for all of the instructor tests (that were configured in CodeRunner).

Table \ref{tab:table1} shows a summary of each problem description along with the number of student solutions for that problem.

\subsection{Design}

We built a Ruby on Rails application to run our experiment \cite{autograder-llm-test-generator}. Each of the 26 problems went through three automated stages:

\begin{enumerate}[1.]
\item Preparing the instructor test suite.
\item Generating an LLM test suite for the problem.
\item Running and evaluating the LLM test suite using the submitted student solutions for the problem.
\end{enumerate}

\begin{figure*}[!t]
\centerline{\includegraphics[width=0.7\textwidth]{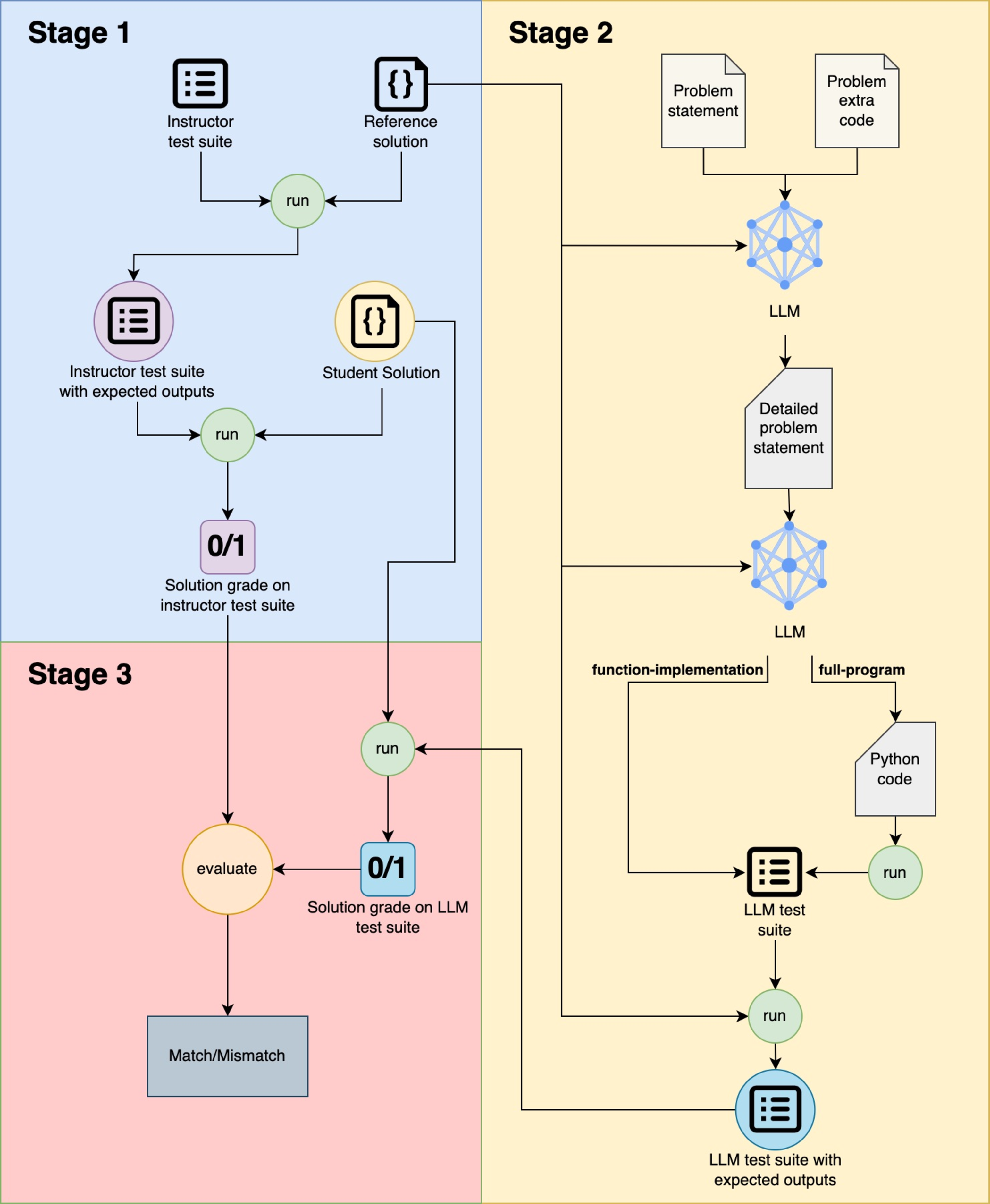}}
\caption{Detailed stages of processing a problem's solution.}
\label{fig:fig1}
\end{figure*}

Figure \ref{fig:fig1} details each of these stages for a given problem. In the first stage, all the problem parts are parsed from a Moodle XML formatted file that is exported from Moodle, the Learning Management System (LMS) course uses. The exported data is stored in a database. This data includes the text of the problem statement (images are stripped out and not stored), its reference solution, any extra code for the problem, and each instructor test along with its expected output. These tests are grouped together under one test suite that is marked as an ``instructor test suite''.

\begin{table*}[!h]%
\caption{The result of running the reference solution of problem 7 on the instructor test suite.}
\label{tab:table2}
\begin{tabular*}{\textwidth}{p{10cm}|p{7cm}}\toprule
\textbf{Raw output} & \textbf{Outputs of each test} \\ \midrule
\multirow{12}{*}{\begin{tabular}[c]{@{}l@{}}\texttt{The words differ by just one character\#\textless{}ab@17943918\#@\textgreater{}\#}\\ \texttt{a: 0}\\ \texttt{b: 1}\\ \texttt{c: 1}\\ \texttt{\#\textless{}ab@17943918\#@\textgreater{}\#}\\ \texttt{The words do not differ by just one character\#\textless{}ab@17943918\#@\textgreater{}\#}\\ \texttt{a: 1}\\ \texttt{b: 0}\\ \texttt{c: 1}\\ \texttt{\#\textless{}ab@17943918\#@\textgreater{}\#}\\ \texttt{The words do not differ by just one character\#\textless{}ab@17943918\#@\textgreater{}\#}\\ \texttt{The words differ by just one character}\end{tabular}} & \textbf{Test 1} \\ \cline{2-2}
& \texttt{The words differ by just one character} \\ \cline{2-2}
& \textbf{Test 2} \\ \cline{2-2}
& \begin{tabular}[c]{@{}l@{}}\texttt{a: 0}\\ \texttt{b: 1}\\ \texttt{c: 1}\end{tabular} \\ \cline{2-2}
& \textbf{Test 3} \\ \cline{2-2}
& \texttt{The words do not differ by just one character} \\ \cline{2-2}
& \textbf{Test 4} \\ \cline{2-2}
& \begin{tabular}[c]{@{}l@{}}\texttt{a: 1}\\ \texttt{b: 0}\\ \texttt{c: 1}\end{tabular} \\ \cline{2-2}
& \textbf{Test 5} \\ \cline{2-2}
& \texttt{The words do not differ by just one character} \\ \cline{2-2}
& \textbf{Test 6} \\ \cline{2-2}
& \texttt{The words differ by just one character} \\
\bottomrule
\end{tabular*}
\end{table*}

To eliminate any future output discrepancies that can be caused by compiler differences, the expected outputs provided in the instructor tests are regenerated by running the reference solution on each test to generate its new expected output.

Once the instructor test suite is ready, student solutions are extracted from a CSV file that is exported from Moodle and run on the instructor test suite. A solution passes a test if the output the solution generates for the test matches the test's expected output. We report the instructor-test-suite grade for each solution as 1 if it passed all the tests in the test suite, or 0 if it failed at least one test in the test suite. Solutions that do not compile have their grades set to -1.

In the second stage, the LLM test suite is generated using two prompts. The first prompt summarizes the problem into a specific format, that summary is fed to the second prompt along with the reference solution to either generate Python scripts that are later run to generate the tests for full-program problems, or to generate a testing script for function-implementation problems that is parsed into individual tests. The reference solution is run on the generated tests to generate their expected outputs. The tests are then grouped together under one test suite that is marked as an ``LLM test suite''. If any test crashes the reference solution, it is automatically marked as rejected and it gets excluded from the LLM test suite. The details of the prompts and some examples are provided in Section \ref{sec:LLMPromptEngineering} on ``LLM Prompt Engineering''.

The third and final stage runs all the student solutions on the LLM test suite. Similar to the instructor test suite, the LLM-test-suite grade for each solution is reported as 1 if it passed all the tests in the test suite, or 0 if it failed at least one test in the test suite. Solutions that do not compile also have their grades set to -1.

\subsection{Running the Solutions}

All three stages require running the reference solution and student solutions on tests to generate expected outputs and student outputs. Running the solutions on the two types of test suites is similar, but there are a few differences.

\subsubsection{Running Solutions on the Instructor Test Suite}

Solutions for full-program problems are run as is for each test. They are provided the test as input through Standard Input. Solutions for function-implementation problems are run differently. Each test is placed inside its own C scope, and the tests are then combined together and separated by output statements that print a predefined separator. The combined tests are substituted in a predefined template along with the reference solution and any extra code for the problem. The resulting code is run and its output is split using the predefined separator to retrieve the individual output of each test. Table \ref{tab:table2} shows an example of the outputs before and after splitting them. The predefined template is provided in Listing \ref{lst:listingb1}, and an example of preparing the reference solution of problem 7 to be run on the problem's instructor test suite is provided in Listing \ref{lst:listingb2}.

\subsubsection{Running Solutions on the LLM Test Suite}

Running solutions for full-program problems on an LLM test suite is identical to running them on an instructor test suite. Though for function-implementation problems, there are three differences: 1) More libraries are included in the template, and 2) the solution to be tested is written into a new file called \texttt{``solution.c''} and it is included in the program. 3) A function call is added before each test to initialize C's pseudo-random number generator algorithm to a value that has been randomly generated beforehand. This ensures that if any random values are being generated in the tests, they will always be the same when the tests are run multiple times. This is essential to allow different valid solutions to produce the same output that was produced by the reference solution.

This random seed value is different for each problem, but is the same for all tests in a problem's LLM test suite. For more thorough testing, it can be a different value for each test as long as the same value is always used for the same test. Table \ref{tab:table3} shows an example of the outputs before and after splitting them. An example of preparing the reference solution of problem 7 to be run on the problem's LLM test suite is provided in Listing \ref{lst:listingb3}.

\begin{table*}[!h]%
\caption{The result of running the reference solution of problem 7 on the LLM test suite (Test 7 truncated for brevity).}
\label{tab:table3}
\begin{tabular*}{\textwidth}{p{8.47cm}|p{8.50cm}}\toprule
\textbf{Raw output} & \textbf{Outputs of each test} \\ \midrule
\multirow{12}{*}{\begin{tabular}[c]{@{}l@{}}0\\ \texttt{\#\textless{}ab@17943918\#@\textgreater{}\#}\\ \texttt{1}\\ \texttt{\#\textless{}ab@17943918\#@\textgreater{}\#}\\ \texttt{0}\\ \texttt{\#\textless{}ab@17943918\#@\textgreater{}\#}\\ \texttt{1}\\ \texttt{\#\textless{}ab@17943918\#@\textgreater{}\#}\\ \texttt{1}\\ \texttt{\#\textless{}ab@17943918\#@\textgreater{}\#}\\ \texttt{1}\\ \texttt{\#\textless{}ab@17943918\#@\textgreater{}\#}\\ Test \texttt{starting}\\ \texttt{Comparing wiue and adqc: 0}\\ \texttt{Test ending}\\ \texttt{Test starting}\\ \texttt{Comparing yauo and jofn: 0}\\ \texttt{Test ending}\\ \texttt{Test starting}\\ \texttt{Comparing ficz and xndn: 0}\\ \texttt{Test ending}\\ \texttt{Test starting}\\ \texttt{Comparing gaub and etwa: 0}\\ \texttt{Test ending}\\ \texttt{Test starting}\\ \texttt{Comparing ypsd and ovli: 0}\\ \texttt{Test ending}\end{tabular}} & \textbf{Test 1} \\ \cline{2-2}
& \texttt{0} \\ \cline{2-2}
& \textbf{Test 2} \\ \cline{2-2}
& \texttt{1} \\ \cline{2-2}
& \textbf{Test 3} \\ \cline{2-2}
& \texttt{0} \\ \cline{2-2}
& \textbf{Test 4} \\ \cline{2-2}
& \texttt{1} \\ \cline{2-2}
& \textbf{Test 5} \\ \cline{2-2}
& \texttt{1} \\ \cline{2-2}
& \textbf{Test 6} \\ \cline{2-2}
& \texttt{1} \\ \cline{2-2}
& \textbf{Test 7} \\ \cline{2-2}
& \begin{tabular}[c]{@{}l@{}}\texttt{Test starting}\\ \texttt{Comparing wiue and adqc: 0}\\ \texttt{Test ending}\\ \texttt{Test starting}\\ \texttt{Comparing yauo and jofn: 0}\\ \texttt{Test ending}\\ \texttt{Test starting}\\ \texttt{Comparing ficz and xndn: 0}\\ \texttt{Test ending}\\ \texttt{Test starting}\\ \texttt{Comparing gaub and etwa: 0}\\ \texttt{Test ending}\\ \texttt{Test starting}\\ \texttt{Comparing ypsd and ovli: 0}\\ \texttt{Test ending}\end{tabular} \\
\bottomrule
\end{tabular*}
\end{table*}

To run a student solution and generate its outputs on the Instructor test suite or the LLM test suite we apply the same methods above using the student solution instead of the reference solution.

\subsection{LLM Prompt Engineering}
\label{sec:LLMPromptEngineering}

We used the \texttt{``gpt-4-0125-preview''} large language model from OpenAI. GPT-4 models show very promising code-generation capabilities according to benchmarks like HumanEval and MBPP \cite{awesome-code-llm}.

The \texttt{``gpt-4-0125-preview''} model was the latest GPT-4 model available at the time of conducting this study. OpenAI describes the model as being ``intended to reduce cases of laziness where the model doesn’t complete a task'' \cite{openai-gpt4-turbo}. Whenever possible, we incorporated a form of self ``Reflexion'' into our prompts. Reflexion is a novel strategy where LLMs are instructed to reflect on the results they produce then revise their responses based on their reflection. It has been shown that Reflexion produces better LLM responses \cite{shinn2024reflexion}. All the prompts were sent to the model using OpenAI's Chat Completions API endpoint.

\subsubsection{Prompt 1: Detailed Problem Statement}

Problem statements can vary in all ways and shapes. Some statements define clear input and output sections, while others do not. Some include explicit time and memory constraints, others do not. The instructor's reference solution may also convey some of the input limits and resource constraints, and it can also show the expected output format. The goal of this prompt is to allow us to represent all problems in a fixed structure that can be used in other prompts in a consistent manner. The LLM receives the problem statement, the reference solution, and any extra code for the problem, and is instructed to produce a summary of the problem under the following sections: ``Scenario'', ``Inputs'', ``Outputs'', ``Example'', and ``Limits''. The LLM is instructed to reflect on the generated sections and make sure they correctly represent the problem and fix any issues it finds in them. Finally the LLM is instructed to organize these sections into a JSON object and respond with it. We use the \texttt{``json\_object''} API response format to receive the JSON object.

An example of the prompt used to generate the Detailed Problem Statement for problem 5 can be found in Appendix \ref{app3}. A Detailed Problem Statement prompt used 923 OpenAI tokens on average.

\subsubsection{Prompt 2: Test Generation}

Test generation prompts are different depending on the problem type, but they share the same essence. A zero-shot approach is used. The LLM is given the detailed problem statement from the result of Prompt 1 and the reference solution, and is instructed to list edge-cases for the problem along with an example on each edge-case. It is then instructed to reflect on the edge-cases to make sure they conform to the detailed problem statement, and fix any issues it finds in them. After reflection, the LLM is instructed to generate a test for each edge-case, in addition to another 100 random tests, if possible. The steps for generating these tests from the edge-cases as well as their final format are different depending on the problem type.

\paragraph{Full-program Test Generation Prompt}

The LLM is instructed to provide a Python script that generates the tests as a JSON array. For each edge-case or random test, the JSON array will contain a JSON object that has one key ``input'', its value is the input of that test. An example of the prompt generated for problem 5 along with its response can be found in Appendix \ref{app4}. Originally, we had the LLM respond directly with a JSON array that contains the inputs and their expected outputs, but it did not always generate the tests correctly, it made mistakes like including comments in the JSON, being lazy i.e., writing a few tests then adding a comment such as ``Write more tests like these...'', and producing wrong expected outputs, even when instructed to generate the expected outputs by running the reference solution and it confirming that it did. It also started to randomly fail to return the JSON array through the API. Instructing the LLM to generate the JSON array through Python code produced more consistent results. A full-program Test Generation prompt used 1,443 OpenAI tokens on average.

\paragraph{Function-implementation Test Generation Prompt}

The LLM is given a predefined code template to add the tests to. If the problem has any extra code, it is added to the template. The LLM has to generate the tests by following very specific instructions. An example of the prompt generated for problem 19 along with its response can be found in Appendix \ref{app4}. Each test must have its own scope and should call the function being tested and print its results, or print the values pointed to by any pointer passed to it. The LLM is instructed to add the edge-cases first. Notice how the LLM reflected on these edge-cases and made changes to them afterwards. After the edge-cases, the random tests are added. Each test is marked with an opening and closing output statement to make parsing individual tests easier later. The random tests are generated as loops that iterate 100 times and call the function with random parameters in each iteration. The tests are later parsed from the script and stored individually. A function-implementation Test Generation prompt used 2,370 OpenAI tokens on average.

\section{Results}\label{sec4}

\subsection{Grading Mismatches}

6,962 of the 33,749 solutions did not compile due to compilation errors. These solutions can be graded as invalid regardless of the tests they will be run on, therefore they were not taken into consideration when we evaluated the performance of the LLM test suite.

Out of the 26,787 compiling solutions, 23,219 (86.7\%) were given matching grades by the instructor test suite and the LLM test suite, i.e., they pass both LLM and instructor test suites or fail both. We reviewed all of the remaining 3,568 solutions that had mismatching grades to understand the types of mistakes that caused these mismatches, and classified them into three types:

\begin{enumerate}[1.]
\item LLM Mismatch: These are mismatches caused by mistakes the LLM made in the LLM test suite.
\item Instructor Mismatch: These are mismatches caused by the instructor test suite not being comprehensive enough to identify an invalid solution.
\item Other Mismatches: These are mismatches that are caused by undefined behaviour in the solution, or ambiguities in the problem statement and the reference solution.
\end{enumerate}

Table \ref{tab:table4} summarizes the number of solutions that fall into each mismatch type for all problems. The left column in each mismatch type shows the number of solutions that were considered valid by the LLM test suite (marked with V) but invalid by the instructor test suite (marked with X). The right column shows the number of solutions that were considered invalid by the LLM test suite but valid by the instructor test suite.

\subsubsection{LLM Mismatch}

1,257 out of the 3,568 (35.2\%) mismatches were LLM Mismatches. The most notable among them are the 363 mismatches for problem 2, the 861 mismatches for problem 16, and the 10 mismatches for problem 10, where the LLM incorrectly graded these solutions as invalid. The mismatches in problems 2 and 16 were caused by invalid tests in each problem's LLM test suite, while the mismatches in problem 10 were due to one of our instructions for the LLM on how to structure the tests.

The remaining 23 LLM Mismatches were because the LLM missed some edge-cases in six different problems, these solutions were incorrectly graded as valid.

\subsubsection{Instructor Mismatch}

1,260 out of the 3,568 (35.3\%) mismatches were Instructor Mismatches. These solutions failed to handle valid tests that were not originally present in the instructor test suite. Some of these valid tests were edge-case tests, others were random tests that produced wrong output. Examples of these mismatches are the 789 solutions in problem 6 that failed to handle edge-cases like the values 0 and 1000 (trailing zeroes). Problem 17 had 131 solutions that failed to capitalize words at the beginning or the end of the sentence, or words that come after multiple spaces or numbers, and problem 28 had 110 solutions that failed to correctly swap the elements of the 2D array when one of them is on the edge of the array.

\begin{table*}[!h]%
\caption{Summary of the number of solutions that fall into each mismatch type for all problems. V: Valid, X: Invalid.\label{tab:table4}}
\begin{tabular*}{\textwidth}{@{\extracolsep\fill}llllllll@{}}\toprule
& &\multicolumn{2}{@{}c@{}}{\textbf{LLM Mismatch}}
&\multicolumn{2}{@{}c@{}}{\textbf{Instructor Mismatch}}
&\multicolumn{2}{@{}c@{}}{\textbf{Other Mismatch}} \\
\cmidrule{3-4}\cmidrule{5-6}\cmidrule{7-8}
\textbf{Problem Number} & \textbf{\# of Solutions} & \begin{tabular}[c]{@{}l@{}} \textbf{LLM: V} \\ \textbf{Instructor: X} \end{tabular} & \begin{tabular}[c]{@{}l@{}} \textbf{LLM: X} \\ \textbf{Instructor: V} \end{tabular} & \begin{tabular}[c]{@{}l@{}} \textbf{LLM: V} \\ \textbf{Instructor: X} \end{tabular} & \begin{tabular}[c]{@{}l@{}} \textbf{LLM: X} \\ \textbf{Instructor: V} \end{tabular} & \begin{tabular}[c]{@{}l@{}} \textbf{LLM: V} \\ \textbf{Instructor: X} \end{tabular} & \begin{tabular}[c]{@{}l@{}} \textbf{LLM: X} \\ \textbf{Instructor: V} \end{tabular} \\
\midrule
1  & 1,870 & 0 & 0   & 0 & 23  & 19 & 0   \\
2  & 841   & 0 & 363 & 0 & 17  & 0  & 0   \\
3  & 892   & 0 & 0   & 0 & 1   & 0  & 5   \\
4  & 687   & 0 & 0   & 0 & 0   & 0  & 0   \\
5  & 460   & 0 & 0   & 0 & 0   & 0  & 2   \\
6  & 1,735 & 0 & 0   & 0 & 789 & 0  & 0   \\
7  & 1,115 & 1 & 0   & 0 & 0   & 1  & 0   \\
8  & 918   & 0 & 0   & 0 & 0   & 2  & 8   \\
9  & 1,425 & 3 & 0   & 0 & 0   & 13 & 14  \\
10 & 1,135 & 4 & 10  & 0 & 19  & 0  & 0   \\
11 & 995   & 0 & 0   & 0 & 0   & 0  & 4   \\
12 & 846   & 0 & 0   & 0 & 26  & 0  & 0   \\
13 & 920   & 0 & 0   & 0 & 0   & 0  & 10  \\
15 & 951   & 0 & 0   & 0 & 0   & 0  & 0   \\
16 & 976   & 0 & 861 & 0 & 0   & 0  & 0   \\
17 & 1,428 & 9 & 0   & 0 & 131 & 0  & 0   \\
18 & 531   & 0 & 0   & 0 & 0   & 0  & 0   \\
19 & 943   & 0 & 0   & 0 & 83  & 0  & 757 \\
20 & 1,058 & 0 & 0   & 0 & 6   & 0  & 4   \\
22 & 690   & 4 & 0   & 0 & 30  & 0  & 1   \\
23 & 1,901 & 0 & 0   & 0 & 8   & 0  & 0   \\
24 & 811   & 0 & 0   & 0 & 4   & 0  & 0   \\
25 & 1111  & 0 & 0   & 0 & 7   & 0  & 0   \\
26 & 620   & 2 & 0   & 0 & 6   & 3  & 62  \\
27 & 696   & 0 & 0   & 0 & 0   & 0  & 144 \\
28 & 1,232 & 0 & 0   & 0 & 110 & 2  & 0 \\
\bottomrule
\end{tabular*}
\end{table*}

\subsubsection{Other Mismatch}

The remaining 1,051 (29.5\%) mismatches were not caused by mistakes on either the LLM or the instructor side. These solutions produced different output from the reference solution because of undefined behaviour in the program on valid tests, or because of ambiguities in the problem statement and the reference solution, which led the LLM to make its own assumptions.

Undefined behaviour causes the program to act in an unpredictable way. It may produce correct output in one run, but incorrect output in a subsequent run. There are a number of ways to detect some of the undefined behaviours that may occur in a program through static analysis and compiler sanitizers, but none were used in this study. Examples of these behaviours for some of the solutions include overflows in data types, or out of bound array accesses.

An example of these mismatches is in problem 19, where 757 solutions produce undefined behaviour when the number to look for is not present in the array.

Given that these mismatches occurred because of valid tests in both the instructor and the LLM test suites, they cannot be considered a mistake on either side, and so were left out from the evaluation of the LLM's efficacy in \textbf{RQ1} and \textbf{RQ2}. These mismatches were instead studied to answer \textbf{RQ3} and find the types of ambiguities problem statements and reference solutions may contain.

Table \ref{tab:table5} displays the summary of the grades for each problem, without solutions that fall into ``Other Mismatches''. The percentages of matches and those of mismatches that fall into both of ``LLM Mismatches'' and ``Instructor Mismatches'' for each problem are shown in Table \ref{tab:table6}.

Next, we highlight the results relevant to each research question.

\begin{table*}[!h]%
\caption{Deciding the validity of a student solution.}
\label{tab:table12}
\begin{tabular*}{\textwidth}{@{\extracolsep\fill}lll@{}}\toprule
\textbf{LLM Test Suite Grade} & \textbf{Instructor Test Suite Grade} & \textbf{Validity} \\
\midrule
1 & 1 & Valid \\
1 & 0 & Mismatch: Manually check solution \\
0 & 1 & Mismatch: Manually check solution \\
0 & 0 & Invalid \\
\bottomrule
\end{tabular*}
\end{table*}

\begin{table*}[!h]%
\caption{Summary of the grades of mismatching error types excluding ``Other Mismatches''. V: Valid, X: Invalid.\label{tab:table5}}
\begin{tabular*}{\textwidth}{@{\extracolsep\fill}llllll@{}}\toprule
& &\multicolumn{2}{@{}c@{}}{\textbf{LLM Mismatch}}
&\multicolumn{2}{@{}c@{}}{\textbf{Instructor Mismatch}} \\
\cmidrule{3-4}\cmidrule{5-6}
\textbf{Problem Number} & \textbf{\# of Solutions} & \begin{tabular}[c]{@{}l@{}} \textbf{LLM: V} \\ \textbf{Instructor: X} \end{tabular} & \begin{tabular}[c]{@{}l@{}} \textbf{LLM: X} \\ \textbf{Instructor: V} \end{tabular} & \begin{tabular}[c]{@{}l@{}} \textbf{LLM: V} \\ \textbf{Instructor: X} \end{tabular} & \begin{tabular}[c]{@{}l@{}} \textbf{LLM: X} \\ \textbf{Instructor: V} \end{tabular} \\
\midrule
1  & 1,851 & 0 & 0   & 0 & 23  \\
2  & 841   & 0 & 363 & 0 & 17  \\
3  & 887   & 0 & 0   & 0 & 1   \\
4  & 687   & 0 & 0   & 0 & 0   \\
5  & 458   & 0 & 0   & 0 & 0   \\
6  & 1,735 & 0 & 0   & 0 & 789 \\
7  & 1,114 & 1 & 0   & 0 & 0   \\
8  & 908   & 0 & 0   & 0 & 0   \\
9  & 1,398 & 3 & 0   & 0 & 0   \\
10 & 1,135 & 4 & 10  & 0 & 19  \\
11 & 991   & 0 & 0   & 0 & 0   \\
12 & 846   & 0 & 0   & 0 & 26  \\
13 & 910   & 0 & 0   & 0 & 0   \\
15 & 951   & 0 & 0   & 0 & 0   \\
16 & 976   & 0 & 861 & 0 & 0   \\
17 & 1,428 & 9 & 0   & 0 & 131 \\
18 & 531   & 0 & 0   & 0 & 0   \\
19 & 186   & 0 & 0   & 0 & 83  \\
20 & 1,054 & 0 & 0   & 0 & 6   \\
22 & 689   & 4 & 0   & 0 & 30  \\
23 & 1,901 & 0 & 0   & 0 & 8   \\
24 & 811   & 0 & 0   & 0 & 4   \\
25 & 1111  & 0 & 0   & 0 & 7   \\
26 & 555   & 2 & 0   & 0 & 6   \\
27 & 552   & 0 & 0   & 0 & 0   \\
28 & 1,230 & 0 & 0   & 0 & 110 \\
\bottomrule
\end{tabular*}
\end{table*}

\subsection{Research Questions}

\paragraph*{RQ1: To what extent do LLM-generated test suites correctly identify valid solutions to CS1 problems?}

To compare the performance of the LLM test suite and the instructor test suite, we determine the validity of a student solution based on the grades it received from both the LLM and the instructor test suites. Table \ref{tab:table12} shows the four possible grades for any solution and the two test suites. If the LLM test suite grade matches the instructor test suite grade for a solution, then the LLM test suite is doing at least as good as the instructor test suite, and that grade decides the validity of the solution. If the two test suites give mismatching grades, it means that one of them is wrong, so we manually review the solution and decide on its validity. We base our evaluation on the grades of the two test suites instead of the instructor test suite grade alone because the instructor test suite may not be comprehensive.

Tables \ref{tab:table7} and \ref{tab:table8} show the confusion matrices of the LLM and the instructor test suites, and Table \ref{tab:table9} shows the statistical metrics for the test suite results based on the confusion matrices. The actual values in the matrices are based on the validity criteria we just established.

To answer \textbf{RQ1} we look at the recall of the test suites. Recall tells us what percentage of valid solutions are identified by the test suites. LLM test suites identified 92.8\% of the valid solutions, compared to 100\% of the valid solutions identified by instructor test suites.

Table \ref{tab:table5} shows that LLM test suites failed to identify all the valid solutions only in three of the 26 problems, with 363 (98.6\%) out of 368 solutions in problem 2, 861 (93\%) out of 1,495 in problem 16, and 10 out of 863 (1.2\%) in problem 10. LLM test suites were able to identify all 15,907 valid solutions for the remaining 23 problems.

\paragraph*{RQ2: How comprehensive are LLM-generated test suites compared to instructor-generated test suites?}

To understand how comprehensive a test suite is, we need to understand how well it identifies invalid solutions.

The False Positive Rate in Table \ref{tab:table9} shows that the LLM test suites only failed to identify 0.2\% of the invalid solutions, compared to 14.6\% by the instructor test suites. The LLM test suites correctly identified all invalid solutions in 20 of the 26 problems with a total of 8,572 out of 8,595 solutions. They failed to identify only 23 solutions in six problems as shown in Table \ref{tab:table5}.

Precision tells us how often test suites are accurate about the solutions they grade as valid. Table \ref{tab:table9} shows that LLM test suites report a precision of 99.8\% compared to 93.1\% by the instructor test suites, indicating they are less likely to grade invalid solutions as valid solutions. The right column under ``Instructor Mismatch'' in Table \ref{tab:table5} shows 1,260 invalid solutions that the instructor test suite failed to identify.

\paragraph*{RQ3: What types of ambiguities can LLM-generated test suites help uncover in problem statements?}

The 1,051 ``Other Mismatches'' were studied to find the types of ambiguities that problem statements may contain and are discussed in the next section.

\section{Discussion}\label{sec5}

\paragraph*{RQ1: To what extent do LLM-generated test suites correctly identify valid solutions to CS1 problems?}

Our results show that LLMs are capable of generating high quality test suites that correctly identify most valid solutions to CS1 problems, but they are not perfect.

Failure to identify the valid solutions in problems 2 and 16 was due to four LLM generated tests that test the solutions against scenarios that are invalid according to the problem statements.

Problem 2 had one invalid edge-case test that sets both the source (value 1) and the destination (value 2) to the same location in the 2D array, the destination overrides the source and the 2D array no longer has the value 1 to mark the source. Depending on how each solution was written, this test could cause some solutions to generate wrong output. Interestingly, the 100 random tests in this problem were written by the LLM in a way that ensures the source and destination are in different locations, which indicates that the LLM might be able to avoid this mistake with better prompting.

\begin{table*}[!h]%
\caption{Performance of LLM Test Suites showing the percentage of matches and mismatches, excluding ``Other Mismatches'', sorted by percentage of matches.\label{tab:table6}}
\begin{tabular*}{\textwidth}{@{\extracolsep\fill}lllll@{}}\toprule
\textbf{Problem Number} & \textbf{\# of Solutions} & \textbf{\% of Matches} & \textbf{\% of LLM Mismatches} & \textbf{\% of Instructor Mismatches} \\
\midrule
11 & 991   & 100.00 & 0.00  & 0.00  \\
15 & 951   & 100.00 & 0.00  & 0.00  \\
13 & 910   & 100.00 & 0.00  & 0.00  \\
8  & 908   & 100.00 & 0.00  & 0.00  \\
4  & 687   & 100.00 & 0.00  & 0.00  \\
27 & 552   & 100.00 & 0.00  & 0.00  \\
18 & 531   & 100.00 & 0.00  & 0.00  \\
5  & 458   & 100.00 & 0.00  & 0.00  \\
7  & 1,114 & 99.91  & 0.09  & 0.00  \\
3  & 887   & 99.89  & 0.00  & 0.11  \\
9  & 1,398 & 99.79  & 0.21  & 0.00  \\
23 & 1,901 & 99.58  & 0.00  & 0.42  \\
24 & 811   & 99.51  & 0.00  & 0.49  \\
20 & 1,054 & 99.43  & 0.00  & 0.57  \\
25 & 1,111 & 99.37  & 0.00  & 0.63  \\
1  & 1,851 & 98.76  & 0.00  & 1.24  \\
26 & 555   & 98.56  & 0.36  & 1.08  \\
10 & 1,135 & 97.09  & 1.23  & 1.67  \\
12 & 846   & 96.93  & 0.00  & 3.07  \\
22 & 689   & 95.07  & 0.58  & 4.35  \\
28 & 1,230 & 91.06  & 0.00  & 8.94  \\
17 & 1,428 & 90.20  & 0.63  & 9.17  \\
19 & 186   & 55.38  & 0.00  & 44.62 \\
2  & 841   & 54.82  & 43.16 & 2.02  \\
6  & 1,735 & 54.52  & 0.00  & 45.48 \\
16 & 976   & 11.78  & 88.22 & 0.00 \\
\bottomrule
\end{tabular*}
\end{table*}

\begin{table*}[!h]%
\caption{Confusion matrix of the LLM Test Suite grades for 25,736 student solutions, excluding solutions in ``Other Mismatches''.}
\label{tab:table7}
\begin{tabular*}{\textwidth}{@{}c@{\extracolsep\fill}c|@{}c@{}c@{}}\toprule
\multicolumn{2}{@{}c@{}}{\multirow{2.5}{*}{n = 25,736 student solutions}} & \multicolumn{2}{@{}c@{}}{\textbf{Actual}} \\ \cmidrule{3-4}
\multicolumn{2}{@{}l@{}}{} & \textbf{Valid} & \textbf{Invalid} \\ \midrule
\multirow{2.5}{*}{LLM Test Suite Grade} & \textbf{Valid} & 15,907 & 23 \\ \cmidrule{2-4}
& \textbf{Invalid} & 1,234 & 8,572 \\ 
\bottomrule
\end{tabular*}
\end{table*}

\begin{table*}[!h]%
\caption{Confusion matrix of the Instructor Test Suite grades for 25,736 student solutions, excluding solutions in ``Other Mismatches''.}
\label{tab:table8}
\begin{tabular*}{\textwidth}{@{}c@{\extracolsep\fill}c|@{}c@{}c@{}}\toprule
\multicolumn{2}{@{}c@{}}{\multirow{2.5}{*}{n = 25,736 student solutions}} & \multicolumn{2}{@{}c@{}}{\textbf{Actual}} \\ \cmidrule{3-4}
\multicolumn{2}{@{}l@{}}{} & \textbf{Valid} & \textbf{Invalid} \\ \midrule
\multirow{2.5}{*}{Instructor Test Suite Grade} & \textbf{Valid} & 17,141 & 1,260 \\ \cmidrule{2-4}
& \textbf{Invalid} & 0 & 7,335 \\ 
\bottomrule
\end{tabular*}
\end{table*}

\begin{table*}[!h]%
\caption{Metrics for the LLM and Instructor Test Suites based on the confusion matrices in Tables \ref{tab:table7} and \ref{tab:table8}.\label{tab:table9}}
\begin{tabular*}{\textwidth}{@{\extracolsep\fill}llll@{}}\toprule
\textbf{Metric} & \textbf{Measures} & \textbf{LLM Test Suite} & \textbf{Instructor Test Suite} \\
\midrule
Accuracy & Percentage of solutions correctly graded & 95.1\% & 95.1\% \\
Precision & Percentage of solutions graded as valid that are actually valid & 99.8\% & 93.1\% \\
Recall & Percentage of valid solutions correctly graded as valid & 92.8\% & 100.0\%  \\
False Positive Rate & Percentage of invalid solutions that were incorrectly graded as valid & 0.2\%  & 14.6\% \\
\bottomrule
\end{tabular*}
\end{table*}

\begin{table*}[!h]%
\caption{Find Max problem.\label{tab:table10}}
\begin{tabular*}{\textwidth}{@{\extracolsep\fill}l@{}}\toprule
Write a function that compares two distinct numbers and prints ``First'' if the first number is larger than the second, or ``Second'' if the second \\
number is larger than the first.
\\
\\
E.g.,
\\
\texttt{findMax(10, 20); // should print ``Second''} \\
\bottomrule
\end{tabular*}
\end{table*}

\begin{table*}[!h]%
\caption{Valid solutions for the Find Max problem in Table \ref{tab:table10}.\label{tab:table11}}
\begin{tabular*}{\textwidth}{@{\extracolsep\fill}p{0.32\textwidth}|p{0.32\textwidth}|p{0.33\textwidth}@{}}\toprule
\begin{tabular}[c]{@{}l@{}}\texttt{void findMax(int first, int second)\{}\\ \hspace{3mm}\texttt{if (first \textgreater second)}\\ \hspace{6mm} \texttt{printf("First");}\\ \hspace{3mm} \texttt{else}\\ \hspace{6mm} \texttt{printf("Second");}\\ \texttt{\}}\end{tabular} & \begin{tabular}[c]{@{}l@{}}\texttt{void findMax(int first, int second)\{}\\   \hspace{3mm}{\texttt{if (first \textgreater{}= second)}}\\ \hspace{6mm}\texttt{printf("First");}\\ \hspace{3mm}\texttt{else}\\ \hspace{6mm}\texttt{printf("Second");}\\ \}\end{tabular} & \begin{tabular}[c]{@{}l@{}}\texttt{void findMax(int first, int second)\{}\\ \hspace{3mm}\texttt{if (first \textgreater second)}\\ \hspace{6mm}\texttt{printf("First");}\\ \hspace{3mm}\texttt{if (first \textless second)}\\     \hspace{6mm}\texttt{printf("Second");}\\ \texttt{\}}\end{tabular} \\
\bottomrule
\end{tabular*}
\end{table*}

Problem 16 had two invalid edge-case tests where the starting index was once negative, and once larger than the ending index. The tests were generated despite the problem statement clearly stating that the indices will be in order and valid. Additionally, the randomly generated tests were generating ranges with a correct starting index, but sometimes with a wrong ending index. This error might be specific to the C programming language because of its usage of the null-terminator at the end of strings. The LLM generated a string with length \texttt{``len + 1''}, and set the character at \texttt{index[len]} to the \texttt{null} terminator. When it generated the ending range, it generated a random number between 0 and len (inclusive), instead of 0 and len - 1, which caused the null terminator to be overwritten in some tests and caused undefined behaviour.

Despite problems 2 and 16 having invalid tests, Table \ref{tab:table6} shows that some solutions were still graded as valid, e.g., in problem 16, around 11.8\% of the problem solutions were graded as valid despite the problem's LLM test suite having two invalid edge-case tests. This occurs because these solutions behave the same way as the reference solution, and therefore produce the same output -- even if it is an output for a broken scenario that should not occur per the problem statement. Think of the following example: Table \ref{tab:table10} describes a Find Max problem. If the LLM generates an invalid test that passes two equal numbers (7 and 7) to the function, then there is a 33.3\% chance that one of the solutions in Table \ref{tab:table11} is graded as valid, depending on how the reference solution is written. All three solutions in Table \ref{tab:table11} solve the problem correctly, but they have three different outputs for the invalid test. The solution on the left will print ``Second'', the solution in the center will print ``First'', and the solution on the right will not print anything. Examples of such student solutions for problems 2 and 16 are provided in Appendix \ref{app4}.

Ten solutions in problem 10 modified the original string passed to the function. The LLM tests would print the original string and the returned value from the function, e.g., \texttt{``Word: Hello, Result: 0''}. Solutions that correctly check if the string is a palindrome but modify the original string inside the function produce different output from the reference solution because strings are passed by reference in C. For example, one solution would copy the string to a new variable then reverse the string in the original variable and compare the two strings for equality, so its output would become: \texttt{``Word: olleH, Result: 0''} which does not match the output of the reference solution, despite the value returned from the function being correct. The LLM is instructed to print the values of any pointers that are passed to the function as can be seen in Table \ref{tab:table6} in point 3 in the system prompt, but the original string that is passed to the function is irrelevant to the result of the function in this problem.

In general, LLMs appear to produce reliable test suites that can correctly identify most valid solutions in CS1 problems, but the highlighted mistakes indicate they may still contain inappropriate tests.

\begin{figure*}[!t]
\centerline{\includegraphics[width=0.7\textwidth]{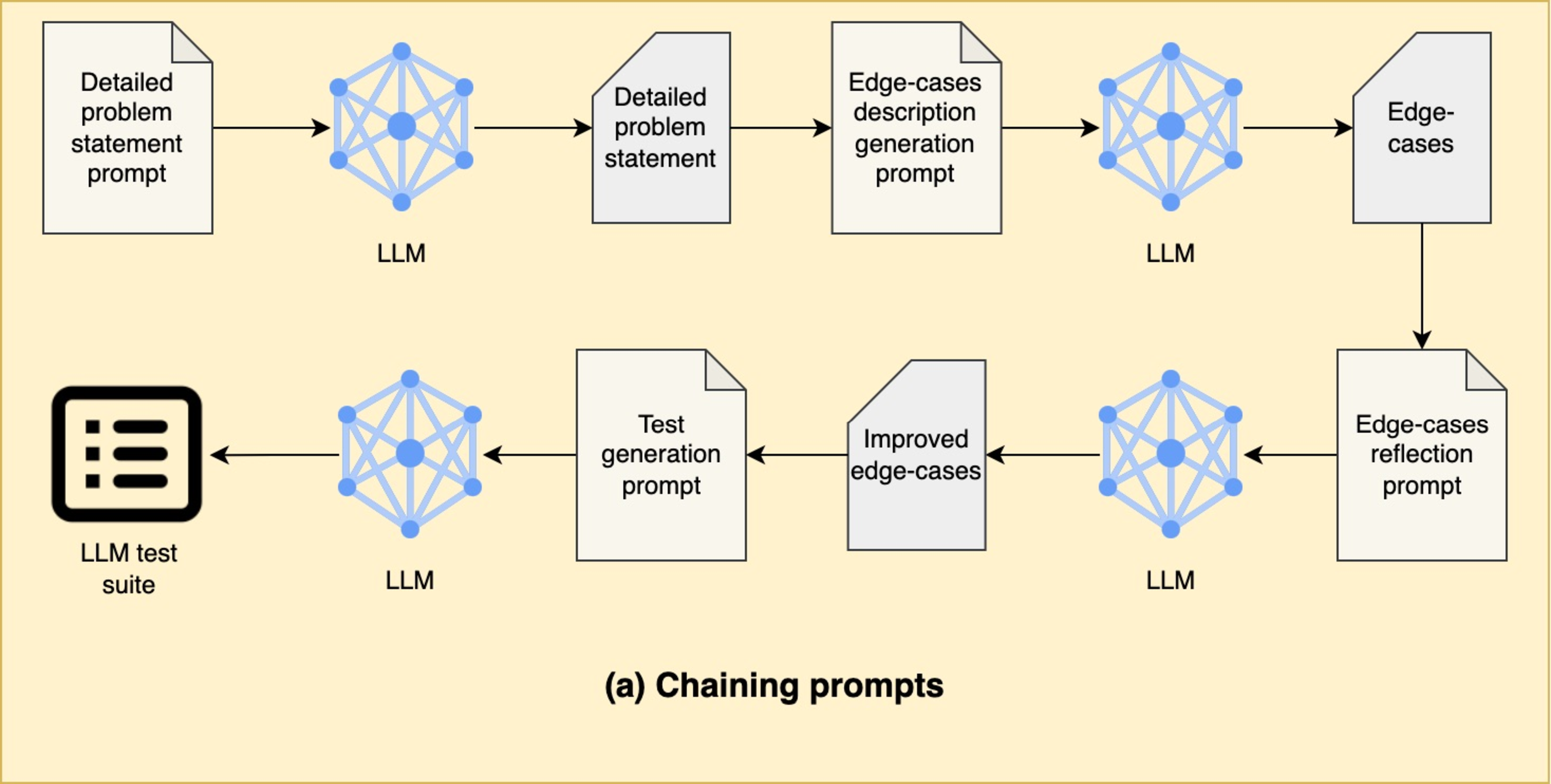}}
\centerline{\includegraphics[width=0.7\textwidth]{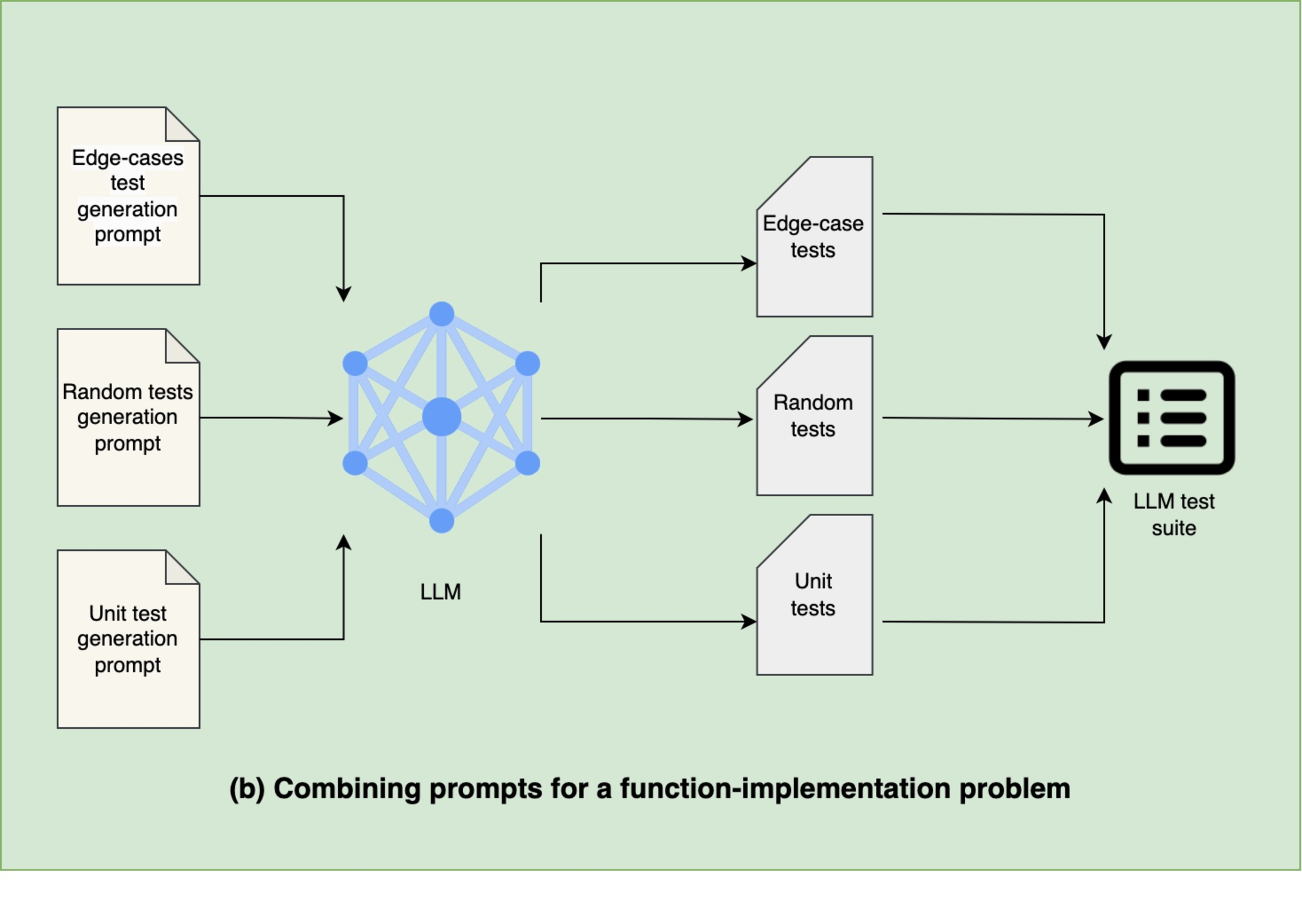}}
\caption{(a) Chaining LLM Prompts and (b) combining results from multiple LLM prompts for a function-implementation problem.}
\label{fig:fig2}
\end{figure*}

Chaining and combining the results of more LLM prompts is likely to produce improved results \cite{wu2022ai} and reduce invalid test cases, but might take more time and incur more cost per problem. On average, it took one minute to generate a problem's LLM test suite, at a cost of approximately 8 British pennies on average (10 US Cents). Figure \ref{fig:fig2} shows one example of how chaining and combining results from multiple LLM prompts can be achieved.

Test structure issues can be fixed by modifying prompt instructions based on the problem type, or by finding a better all-around test structure. For example, modifying the prompt to have the LLM print the values of passed pointers only when they are used to store results relevant to the problem might solve the issue in problem 10.

\paragraph*{RQ2: How comprehensive are LLM-generated test suites compared to instructor-generated test suites?}

The left column under ``LLM Mismatch'' in Table \ref{tab:table5} shows the number of invalid solutions that the LLM did not correctly identify for each problem. These failures can be classified into three categories:

\begin{enumerate}[1.]

\item \textbf{Bad coverage:} This type of failure can be attributed to the LLM model's capabilities. LLM models with better generative capabilities can generate tests that have better coverage of edge-cases. For example, three solutions in problem 10 do not correctly handle a palindrome of only two letters, the LLM test suite generated tests that test even-sized palindromes but it did not have a test for size 2 specifically, which was not correctly handled in some student solutions. Other examples of this type of failure are solutions in problem 17 that should fail because they capitalize letters that follow punctuation marks, or fail to capitalize a one letter string or words that come after more than just one space. Another example is problem 22 where three solutions do not handle ties in the largest-distance as required. The LLM test suite did not provide tests that target these edge-case scenarios, which caused the solutions to be considered valid, when they should not have been.

\item \textbf{Solution-specific failures:} These are failures that can only be identified with very specific tests that depend on the solution. In problem 7, one solution would fail on a very specific test where the hamming distance is 1 and one letter in the first string appears more than once in the second string. The three solutions in problem 9 should fail because they do not handle steps that are outside specific ranges defined in the solutions. One solution in problem 10 sums up the ASCII code of the letters in each half and compares the sums, so it should fail on tests like ``abab''. Another solution in problem 22 does not compare doubles correctly and would fail on a test that has doubles with more than one decimal point. Instructors count on random tests to catch scenarios like these. More experienced instructors are familiar with the types of solutions to expect from students for a problem, and therefore include specific tests to fail these solutions.

\item \textbf{Test structure:} Similar to the issues in the 10 solutions in Problem 10 described in the previous subsection, these failures occurred due to the way the LLM was instructed to form the tests. Two solutions in problem 26 print part of the answer directly to output and set the other part of the answer to the string passed to the function, which gets printed in the main function afterwards. The overall output appears complete and the solution is incorrectly considered valid because of this. Instructing the LLM to include one small unit test to the test suite that directly checks the value of the passed string after the function should resolve this issue. This can be generalized to instructing the LLM to write one unit test that checks the return value of non void functions, in addition to checking the values of variables that are passed to the function through pointers or by reference when appropriate. This can be achieved using the same prompt or through an additional prompt.

\end{enumerate}

The right column under ``Instructor Mismatch'' in Table \ref{tab:table5} shows the numbers of invalid solutions that the instructor test suites failed to identify, but the LLM test suites did. These mistakes in the instructor test suite can be summarized as:

\begin{enumerate}[1.]

\item \textbf{Missing edge-cases:} like the value 0 in problem 1, and the maximum value of an integer in problem 23.

\item \textbf{Failure to vary the inputs:} like problem 19 where the instructor test suite only checks for the number -1 in the array, which caused all solutions that hard-coded searching for -1 to be incorrectly considered valid.

\item \textbf{Not detecting wrong data types or bad array sizes:} like problem 12 where the reference solution uses the double data type and the problem statement explains how to use doubles, but solutions that use the float data type pass the instructor test suite. Another example is problem 23 where some solutions store the results of the integer to binary string conversion in an array that is shorter than 32 in size but still pass the instructor test suite.

\item \textbf{Defining very large strings and only using parts of them:} like in problem 10 where the instructor test suite defines a 100 character string, but fills it with a few letters only. This does not catch all index out of bound errors because the accessed indices are still likely to fall in locations that are reserved for the string.

\item \textbf{Not enough random tests:} A large number of random tests increases the likelihood that solutions with implementation-specific issues fail.

\end{enumerate}

The results per-problem indicate that LLMs generate test suites that are as comprehensive as instructor test suites most of the time, and oftentimes do better. Chaining and combining prompt results could further improve these results.

\paragraph*{RQ3: What types of ambiguities can LLM-generated test suites help uncover in problem statements?}

We looked at all 1,051 solutions that caused an ``Other Mismatch'' and identified mismatches that occur because of a solution's handling of ambiguous scenarios in tests. The LLM test suite helped identify the following ambiguities in the problems:

\begin{enumerate}[1.]

\item \textbf{No explicit mention of input limits:} which leaves these limits open to interpretation by the LLM. This is important to pay attention to when generating LLM test suites, as oftentimes in CS1 courses, instructors may not want to introduce specific concepts like overflow-errors in the first few weeks of class, so making sure the LLM knows they do not want tests that overflow student solutions is important, and that can be achieved by specifying the input limits. For example, in problem 1 some solutions run into integer overflow errors because they use factorials to find the combination and the LLM test suite tests large values that cause the factorial calculations to overflow. The same happens in problem 9 when some solutions overflow the char data type.

\item \textbf{No instructions on how to handle negative outcomes:} like values not being found in arrays (problem 19) or empty arrays (problems 20, 22, and 26). If the problem statement does not specify what should be done when no answer can be found, students might not consider such scenarios, leading to undefined behaviours in their code. Alternatively, they might come up with their own way to handle these tests, which could result in invalid solutions.

\end{enumerate}

LLMs tend to assume limits when they are not explicit in problem statements. The tests generated by these assumptions may or may not be what the instructor is looking for. This can be helpful to instructors as they think about how students may approach their problems and what kinds of assumptions they will make, and whether they want them to make these assumptions or rather explicitly include them in the problem statement.

\subsection{Implications}

The findings of this study suggest several important implications for computing education, including both practical classroom applications and broader considerations regarding the use of generative AI and LLMs.

In a classroom setting, the methods described in this study to generate LLM test suites can be implemented as a black-box tool. This tool could accept a problem statement as text, any extra code to be provided to students, and a reference solution, and it would then produce an LLM-generated test suite. Our empirical findings suggest that LLM-generated test suites can be as comprehensive as instructor-generated ones, often identifying edge-cases that instructors might overlook. By lowering the barriers associated with creating comprehensive test suites, LLM-based tools could encourage more instructors to adopt autograding systems, providing students with the benefits of immediate feedback \cite{mitra2023studying}.

Our findings also contribute to the broader understanding of how generative AI and LLMs can be integrated into computing education contexts. We have seen that LLMs have the potential to reduce instructor workload by automating the generation of learning resources, similar to prior work \cite{sarsa2022automatic,doughty2024comparative}. However, instructors still play a critical role in reviewing and refining LLM-generated resources to ensure their quality and appropriateness. This aligns with recent theoretical work in the area of learnersourcing, which emphasizes the importance of human review and iterative refinement of AI-generated outputs to maintain educational integrity \cite{khosravi2023learnersourcing}.

\subsection{Limitations and Recommendations}

There are limitations to this study. We removed all images from the problem statements before passing them to the LLM. Sometimes problem statements may contain images that are essential to the understanding of the problem and that may affect the generated tests positively or negatively depending on the LLM's capabilities. With the growing number of GenAI technologies that are now capable of accepting and processing multi-modal data such as images, we recommend this as a direction for future research.

Secondly, LLM results are not deterministic, and therefore studies on LLMs may suffer from replicability difficulties. OpenAI's API accepts a ``temperature'' value between 0 and 2 that controls the randomness of the results of its Chat Completion endpoint. We used a lower temperature value of 0.2 which makes the model outputs more focused and deterministic \cite{openai-temperature}. Moreover, our study only looked at the C programming language; other languages may produce different results based on the language capabilities of the LLM, which can depend on the availability of training data. Evaluating the performance of test generation on different languages is another opportunity that researchers can explore and may require designing different prompts and different strategies to run and grade the student solutions. We provide examples for prompting and executing Python3 programs \cite{llm-system-prompts, python3-unit-test-runner}.

Thirdly, the LLM seemed to provide very good descriptions of the edge test cases it generated (see Appendix \ref{app3}). Future research can study how LLMs may be prompted to generate test suites with test descriptions or test-failure messages in order to provide better feedback to students about the tests their code may fail. Such enhanced feedback could help in avoiding the issues related to providing only binary feedback, such as lowering student engagement and encouraging cheating \cite{prather2018metacognitive}.

Finally, there is a security risk when running code generated by LLMs. Most autograders run code in sandboxes, but it is still recommended that instructors review the tests generated for any security concerns.

\section{Conclusion}\label{sec6}

This study evaluated the effectiveness of using Large Language Models (LLMs) to automatically generate test suites for CS1 programming problems. Automating test generation in Computing Education can save instructors significant time and effort, allowing more focus on instructional design while providing students with the benefits of immediate feedback.

We used OpenAI's GPT-4 model to generate test suites for 26 CS1 problems. We then compared student solutions evaluated by these LLM-generated test suites with those evaluated by instructor-created test suites, and we categorized mismatches into three types: errors due to the LLM-generated tests, limitations in the instructor test suite, and ambiguities inherent in the problems. LLM-generated errors included invalid tests, missed edge-cases, and failures to handle certain specific scenarios. Errors in the instructor test suite were often due to a lack of comprehensive coverage, such as missing edge-cases, inconsistent data type validation, and insufficient randomization. Finally, ambiguities in the problems themselves -- -- like undefined input limits or a lack of guidance for handling certain outcomes -- led to inconsistencies in how solutions were evaluated.

Our findings demonstrate that LLMs can reliably create test suites that identify valid solutions with high accuracy, often matching or surpassing typical instructor test suites in detecting invalid solutions. While LLM-generated test suites are promising for reducing workload, some instructor oversight is still recommended. Additionally, the use of LLMs for test generation can assist instructors in refining problem statements by highlighting ambiguities and edge-cases, ultimately leading to more effective instructional design.

\section*{Conflict of interest}

The authors declare no potential conflict of interests.

\bibliographystyle{ACM-Reference-Format}
\bibliography{sample-base}

\onecolumn

\appendix

\renewcommand{\thelstlisting}{\Alph{section}\arabic{lstlisting}}

\section{Problem Samples\label{app1}}

We provide examples for each problem type in this appendix.

\subsection{Full-program Problem Samples\label{app1.1}}

Problems 5, 11, and 24 are examples of full-program problems.

\subsubsection{Problem 5\label{app1.1.1}}

\noindent Write a program that prompts the user to enter a letter of the alphabet.  The program should then display the next letter that follows the input letter.  The output should also include the original letter and a stylized arrow. \\

\noindent NOTE: You can assume that the input will not be 'z' or 'Z' (although, you should test your program to see what happens in these cases!)

\begin{lstlisting}[caption={The reference solution for problem 5.\label{lst:listinga1}},basicstyle=\fontsize{8}{10}\selectfont\ttfamily,frame=single,breaklines=true,breakindent=0pt,breakatwhitespace=true,
  postbreak=\mbox{\textcolor{red}{$\hookrightarrow$}\space},]
#define _CRT_SECURE_NO_WARNINGS
#include <stdio.h>

int main(void) {
  char input;
  char result;
  printf("Enter letter:\n");
  scanf("%c", & input);
  result = (char)(input + 1);
  printf("%c -> %c", input, result);
  return 0;
}
\end{lstlisting}

\subsubsection{Problem 11\label{app1.1.2}}

\noindent Write a program that prompts the user to enter an integer, and then displays an equation as output which illustrates the input value being multiplied by two.

\begin{lstlisting}[caption={The reference solution for problem 11.\label{lst:listinga2}},basicstyle=\fontsize{8}{10}\selectfont\ttfamily,frame=single,breaklines=true,breakindent=0pt,breakatwhitespace=true,
  postbreak=\mbox{\textcolor{red}{$\hookrightarrow$}\space},]
#define _CRT_SECURE_NO_WARNINGS
#include <stdio.h>

int main(void) {
  int input;
  printf("Enter value:\n");
  scanf("%d", & input);
  printf("%d x 2 = %d", input, 2 * input);
  return 0;
}
\end{lstlisting}

\clearpage

\subsubsection{Problem 24\label{app1.1.3}}

\noindent You have been hired by a shipping company that ships items to customers using containers of two different sizes. People seem to be doing a lot more shopping from home these days, so business is booming. The containers are either large or small: \\

\noindent When an order for items is received, all of the requested items are produced. Then the company works out how many large and small containers will be needed to ship the items. The following rules are followed:

\begin{enumerate}[1.]
\item Use as many large containers as possible, to minimise the total number of containers
\item Do not ship any containers which are not completely full
\end{enumerate}

\noindent This second rule means that any left over items, which could only be placed into a partially full container, will be scrapped.

\noindent For example, let’s say that an order for 31 items is placed. Also, assume that the capacity of a large container is 8 and the capacity of a small container is 3. In this case, as illustrated below, a total of 3 large containers and 2 small containers would be needed, and there would be 1 left over item that is scrapped. \\

\noindent Write a program which prompts the user to enter three values:
\begin{enumerate}[1.]
\item The capacity of a large container (i.e. how many items can be packed in a large container)
\item The capacity of a small container (i.e. how many items can be packed in a small container)
\item The total number of items in the order
\end{enumerate}

\noindent Your program should calculate and display the total number of containers, of each size, that will be needed to ship the order. In addition, you must display the number of items that will be scrapped (because they can’t be placed into a full container). You can assume that the capacity of the large container is greater than the capacity of the small container, and you can assume the capacity of the small container is greater than 0. \\

\noindent The program output below shows how the prompts and the output for your program should be formatted, given the inputs in the above example. In the output below, user input is in red and bold: \\

\noindent Large capacity:

\color{red}
\noindent 8
\color{black}

\noindent Small capacity:

\color{red}
\noindent 3
\color{black}

\noindent Number of items:

\color{red}
\noindent 31 \\
\color{black}

\noindent Allocated:

\noindent - Large: 3
 
\noindent - Small: 2
 
\noindent - Scrap: 1 \\

\noindent Notes:

\noindent * take care with spacing and formatting

\noindent * there should be a newline printed after each prompt to the user (i.e. asking for the capacities and number of items)

\clearpage

\begin{lstlisting}[caption={The reference solution for problem 24.\label{lst:listinga3}},,basicstyle=\fontsize{8}{10}\selectfont\ttfamily,frame=single,breaklines=true,breakindent=0pt,breakatwhitespace=true,
  postbreak=\mbox{\textcolor{red}{$\hookrightarrow$}\space},]
#define _CRT_SECURE_NO_WARNINGS
#include <stdio.h>

int main(void) {
  int large, small, items, numLarge, numSmall;
  printf("Large capacity:\n");
  scanf("%d", & large);
  printf("Small capacity:\n");
  scanf("%d", & small);
  printf("Number of items:\n");
  scanf("%d", & items);
  numLarge = items / large;
  items = items % large;
  numSmall = items / small;
  items = items % small;
  printf("Allocated:\n - Large: %d\n - Small: %d\n - Scrap: %d", numLarge, numSmall, items);
  return 0;
}
\end{lstlisting}

\subsection{Function-implementation Problem Samples\label{app1.2}}

Problems 2 and 7 are examples of function-implementation problems.

\subsubsection{Problem 2\label{app1.2.1}}

\noindent [This is a challenge exercise - good practice if you would like a more difficult exercise to try, however it is only worth 0.25 marks out of the 10 marks for this lab; that is the equivalent of 0.05\% contribution towards your final grade... good luck!] \\

\noindent You have been hired by a warehousing company that stores enormous quantities of boxes in a huge warehouse:

\noindent The boxes frequently need to be moved around on heavy pellets from one location to another. However, the manager has noticed that the pellets are not always moving in a very efficient manner. This lack of efficiency is eating into the bottom line and the shareholders are demanding action. To improve efficiency in the warehouse, the manager wants software to calculate the most efficient path along which a pellet should move to its destination. Given the location of a pellet, and the location of where it should move to, the software should compute a direct route from the pellet to the destination. This computed route is then sent remotely to the workers on the warehouse floor who will move the pellet.  Pellets can only move horizontally and vertically (from a bird's eye view of the warehouse floor). \\

\noindent There is some skepticism amongst workers that the long term plan of this remote routing project is to replace them with robots, however your job is just to write the software.

\noindent For this task you must define a function called PelletRoute(). This function takes just one input, a 2D array of integers (10 rows and 10 columns) which represents the layout of the warehouse floor. Every element in the 2D array will be equal to 0 (to represent empty space) with the exception of two values: \\

\noindent * the value 1 will appear exactly once and represents the location of the pellet

\noindent * the value 2 will appear exactly once and represents the location of the destination \\

\noindent The function must calculate a direct route from the pellet to the destination using this algorithm:

\begin{enumerate}[1.]
    \item move the pellet horizontally (left or right) if necessary, until it lines up with the destination
    \item move the pellet vertically (up or down) if necessary, until it reaches the destination
\end{enumerate}

\noindent The route that you calculate should be indicated by setting all array elements on the route to the value 3. Note, as implied by the algorithm above, the pellet must move horizontally first before moving vertically. For example, the diagram on the left below shows an initial input to the function. The diagram on the right shows how the values in the array should be updated after the function has finished executing:

\noindent The function prototype declaration is:

void PelletRoute(int warehouse[10][10]) \\

\noindent The input array is a 2D array of integers consisting of 10 rows and 10 columns. Every element in the array will be equal to zero except for two values: there will be a single “1” (representing the worker) and a single “2” (representing the box). Good luck!

\clearpage

\begin{lstlisting}[caption={The reference solution for problem 2.\label{lst:listinga4}},basicstyle=\fontsize{8}{10}\selectfont\ttfamily,frame=single,breaklines=true,breakindent=0pt,breakatwhitespace=true,
  postbreak=\mbox{\textcolor{red}{$\hookrightarrow$}\space},]
int Min(int a, int b) {
  if (a < b) {
    return a;
  } else {
    return b;
  }
}
int Max(int a, int b) {
  if (a > b) {
    return a;
  } else {
    return b;
  }
}
void PelletRoute(int warehouse[10][10]) {
  int i, j;
  int srow, scol;
  int erow, ecol;
  int startIndex, endIndex;
  // Locate positions
  srow = 0;
  scol = 0;
  erow = 0;
  ecol = 0;
  for (i = 0; i < 10; i++) {
    for (j = 0; j < 10; j++) {
      if (warehouse[i][j] == 1) {
        srow = i;
        scol = j;
      } else if (warehouse[i][j] == 2) {
        erow = i;
        ecol = j;
      }
    }
  }
  // Draw horizontal lines
  startIndex = Min(scol, ecol);
  endIndex = Max(scol, ecol);
  for (i = startIndex; i <= endIndex; i++) {
    if (warehouse[srow][i] == 0) {
      warehouse[srow][i] = 3;
    }
  }
  // Draw vertical lines
  startIndex = Min(srow, erow);
  endIndex = Max(srow, erow);
  for (i = startIndex; i <= endIndex; i++) {
    if (warehouse[i][ecol] == 0) {
      warehouse[i][ecol] = 3;
    }
  }
}
\end{lstlisting}

\subsubsection{Problem 7\label{app1.2.2}}

\noindent Carefully review the ``word chain'' game, as described in the Lab 11 preparation document and illustrated in class: \\

\noindent Define the following function:

int OneLetterDifference(char *word1, char *word2) \\

\noindent This function takes two words as inputs. You can assume that both words will have come from a word list containing words of length 4, and therefore both words will be of length 4. This function must compare each word – character by character – and return true (i.e. 1) only if there is exactly one character that differs between the two words. Apart from this one character, all other characters must be identical and in the same positions. \\

\noindent Note: The constant WORD\_LENGTH is defined to be 5 (this provides space for the characters in the word and the null terminating character).

\begin{lstlisting}[caption={The extra code for problem 7.\label{lst:listinga5}},basicstyle=\fontsize{8}{10}\selectfont\ttfamily,frame=single,breaklines=true,breakindent=0pt,breakatwhitespace=true,
  postbreak=\mbox{\textcolor{red}{$\hookrightarrow$}\space},]
#define WORD_LENGTH 5
\end{lstlisting}

\begin{lstlisting}[caption={The reference solution for problem 7.\label{lst:listinga6}},basicstyle=\fontsize{8}{10}\selectfont\ttfamily,frame=single,breaklines=true,breakindent=0pt,breakatwhitespace=true,
  postbreak=\mbox{\textcolor{red}{$\hookrightarrow$}\space},]
int OneLetterDifference(char * word1, char * word2) {
  int i = 0;
  int diff = 0;
  while (word1[i] != '\0') {
    if (word1[i] != word2[i]) {
      diff++;
    }
    i++;
  }
  return diff == 1;
}
\end{lstlisting}

\clearpage
\section{Running Solutions on Test Suites\label{app2}}

\setcounter{lstlisting}{0}

\begin{lstlisting}[caption={Template to run function-implementation problems.\label{lst:listingb1}},basicstyle=\fontsize{8}{10}\selectfont\ttfamily,frame=single,breaklines=true,breakindent=0pt,breakatwhitespace=true,
  postbreak=\mbox{\textcolor{red}{$\hookrightarrow$}\space},]
#include <stdio.h>
#include <stdlib.h>
#include <ctype.h>
#include <string.h>
#include <stdbool.h>
#include <math.h>

{{ EXTRA_CODE_GOES_HERE }}

{{ REFERENCE_SOLUTION_GOES_HERE }}

int main() {
  {{ TEST_CASES_GO_HERE }}
  return 0;
}
\end{lstlisting}

\begin{lstlisting}[caption={The reference solution of Problem 7 and its instructor tests substituted in the template in Listing \ref{lst:listingb1}.\label{lst:listingb2}},basicstyle=\fontsize{8}{10}\selectfont\ttfamily,frame=single,breaklines=true,breakindent=0pt,breakatwhitespace=true,
  postbreak=\mbox{\textcolor{red}{$\hookrightarrow$}\space},]
#include <stdio.h>
#include <stdlib.h>
#include <ctype.h>
#include <string.h>
#include <stdbool.h>
#include <math.h>

#define WORD_LENGTH 5

int OneLetterDifference(char * word1, char * word2) {
  int i;
  int count = 0;
  for (i = 0; i < 5; i++) {
    if (word1[i] != word2[i]) {
      count++;
    }
  }
  if (count == 1) {
    return 1;
  } else return 0;
}

int main() {
  {
    char word1[WORD_LENGTH] = "able";
    char word2[WORD_LENGTH] = "ably";

    if (OneLetterDifference(word1, word2)) {
      printf("The words differ by just one character");
    } else {
      printf("The words do not differ by just one character");
    }
  }


  printf("#<ab@17943918#@>#\n");
  {
    char word1[WORD_LENGTH] = "rear";
    char word2[WORD_LENGTH] = "beat";
    char word3[WORD_LENGTH] = "bear";

    printf("a: %d\n", OneLetterDifference(word1, word2));
    printf("b: %d\n", OneLetterDifference(word1, word3));
    printf("c: %d\n", OneLetterDifference(word2, word3));
  }

  printf("#<ab@17943918#@>#\n");
  {
    char word1[WORD_LENGTH] = "able";
    char word2[WORD_LENGTH] = "abet";

    if (OneLetterDifference(word1, word2)) {
      printf("The words differ by just one character");
    } else {
      printf("The words do not differ by just one character");
    }
  }

  printf("#<ab@17943918#@>#\n");
  {
    char word1[WORD_LENGTH] = "best";
    char word2[WORD_LENGTH] = "beat";
    char word3[WORD_LENGTH] = "bear";

    printf("a: %d\n", OneLetterDifference(word1, word2));
    printf("b: %d\n", OneLetterDifference(word1, word3));
    printf("c: %d\n", OneLetterDifference(word2, word3));
  }

  printf("#<ab@17943918#@>#\n");
  {
    char word1[WORD_LENGTH] = "aaaa";
    char word2[WORD_LENGTH] = "aaaa";

    if (OneLetterDifference(word1, word2)) {
      printf("The words differ by just one character");
    } else {
      printf("The words do not differ by just one character");
    }
  }

  printf("#<ab@17943918#@>#\n");
  {
    char word1[WORD_LENGTH] = "aaaa";
    char word2[WORD_LENGTH] = "aaba";

    if (OneLetterDifference(word1, word2)) {
      printf("The words differ by just one character");
    } else {
      printf("The words do not differ by just one character");
    }
  }
  return 0;
}
\end{lstlisting}

\begin{lstlisting}[caption={Code to run Problem 7's reference solution on the LLM test suite. The seed value here is 761177235.\label{lst:listingb3}},basicstyle=\fontsize{8}{10}\selectfont\ttfamily,frame=single,breaklines=true,breakindent=0pt,breakatwhitespace=true,
  postbreak=\mbox{\textcolor{red}{$\hookrightarrow$}\space},]
#include <stdio.h>
#include <ctype.h>
#include <string.h>
#include <stdbool.h>
#include <math.h>
#include <limits.h>
#include <time.h>
#include <stdlib.h>

#define WORD_LENGTH 5

#include "solution.c"

int main() {
  srand(761177235);
  {
    char word1[WORD_LENGTH] = "test";
    char word2[WORD_LENGTH] = "test";
    int result = OneLetterDifference(word1, word2);
    printf("%d\n", result);
  }

  printf("#<ab@17943918#@>#\n");
  srand(761177235);
  {
    char word1[WORD_LENGTH] = "bake";
    char word2[WORD_LENGTH] = "cake";
    int result = OneLetterDifference(word1, word2);
    printf("%d\n", result);
  }

  printf("#<ab@17943918#@>#\n");
  srand(761177235);
  {
    char word1[WORD_LENGTH] = "abcd";
    char word2[WORD_LENGTH] = "wxyz";
    int result = OneLetterDifference(word1, word2);
    printf("%d\n", result);
  }

  printf("#<ab@17943918#@>#\n");
  srand(761177235);
  {
    char word1[WORD_LENGTH] = "bake";
    char word2[WORD_LENGTH] = "bale";
    int result = OneLetterDifference(word1, word2);
    printf("%d\n", result);
  }

  printf("#<ab@17943918#@>#\n");
  srand(761177235);
  {
    char word1[WORD_LENGTH] = "bake";
    char word2[WORD_LENGTH] = "bako";
    int result = OneLetterDifference(word1, word2);
    printf("%d\n", result);
  }

  printf("#<ab@17943918#@>#\n");
  srand(761177235);
  {
    char word1[WORD_LENGTH] = "lake";
    char word2[WORD_LENGTH] = "bake";
    int result = OneLetterDifference(word1, word2);
    printf("%d\n", result);
  }

  printf("#<ab@17943918#@>#\n");
  srand(761177235);
  for (int i = 0; i < 100; i++) {
    char word1[WORD_LENGTH];
    char word2[WORD_LENGTH];
    for (int j = 0; j < WORD_LENGTH - 1; j++) {
      word1[j] = 'a' + rand() % 26;
      word2[j] = 'a' + rand() % 26;
    }
    word1[WORD_LENGTH - 1] = '\0';
    word2[WORD_LENGTH - 1] = '\0';

    printf("Test starting\n");
    {
      int result = OneLetterDifference(word1, word2);
      printf("Comparing %s and %s: %d\n", word1, word2, result);
    }
    printf("Test ending\n");
  }

  return 0;
}
\end{lstlisting}

\clearpage

\section{LLM Prompt Examples\label{app3}}

\setcounter{lstlisting}{0}

\subsection{Detailed Problem Statement Prompt\label{app3.1}}

\begin{lstlisting}[frame=single,breaklines=true,breakindent=0pt,breakatwhitespace=true,caption={System Prompt.\label{lst:listingc1}}]
Here is a step-by-step guide that you should follow to answer me.

You are going to receive a programming problem description along with its optimal solution and any extra code required to solve the problem. You are required to respond with a detailed problem statement based on the provided description, solution, and extra code if any. The detailed problem statement must only have the following 5 sections: The problem scenario, the problem inputs, the problem outputs, an example, and a limits section that contains any memory limits, time limits, or code complexity limits that are defined in the problem description or can be inferred from the optimal solution. If you cannot generate an example, just fill the Example section with "None". If there are no limits defined in the problem description and you could not infer the limits from the optimal solution, just fill the section with "Not Available".

After you provide the detailed problem statement, please reflect on it to make sure the sections in it correctly represent the problem. If there are any mistakes, point them out.

After you finish the reflection, please take into account the feedback from the reflection, and fix any issues in the detailed problem statement that you generated.

Finally, use the following format to print the final improved detailed problem statement. If the feedback did not improve the detailed problem statement, print the detailed problem statement anyway.

Here is the format:
Print a native JSON object that only contains the following keys: "scenario", "inputs", "outputs", "example", "limits". The value of each key must be a string that corresponds to the appropriate value from the detailed problem statement. The JSON object must not have any comments in it.
\end{lstlisting}

\begin{lstlisting}[frame=single,breaklines=true,breakindent=0pt,breakatwhitespace=true,caption={User Prompt.\label{lst:listingc2}}]
Problem Statement:
Write a program that prompts the user to enter a letter of the alphabet.  The program should then display the next letter that follows the input letter.  The output should also include the original letter and a stylized arrow.

NOTE: You can assume that the input will not be 'z' or 'Z' (although, you should test your program to see what happens in these cases!)

Optimal Solution:
```
#define _CRT_SECURE_NO_WARNINGS
#include <stdio.h>

int main(void)
{
  char input;
  char result;

  printf("Enter letter:\n");
  scanf("%c", &input);

  result = (char)(input + 1);

   printf("%c -> %c", input, result);

  return 0;
}
```

Extra Code Section:
```

```
\end{lstlisting}

\begin{lstlisting}[frame=single,breaklines=true,breakindent=0pt,breakatwhitespace=true,caption={LLM Response.\label{lst:listingc3}}]
{
  "scenario": "The program is designed to take a single letter from the alphabet as input from the user and display the next letter in the alphabet sequence. The output format includes the original letter entered by the user, followed by a stylized arrow '->', and then the next letter in the alphabet. This program is intended to help users quickly find the subsequent letter of any given letter in the alphabet, with the exception that the input should not be 'z' or 'Z'.",
  "inputs": "A single letter of the alphabet (excluding 'z' or 'Z') entered by the user.",
  "outputs": "The original letter entered by the user, followed by a stylized arrow '->', and then the next letter in the alphabet.",
  "example": "Input: A\nOutput: A -> B",
  "limits": "Not Available"
}
\end{lstlisting}

\subsection{Full-program Problem Prompt\label{app3.2}}

\begin{lstlisting}[frame=single,breaklines=true,breakindent=0pt,breakatwhitespace=true,caption={System Prompt.\label{lst:listingc4}}]
Here is a step-by-step guide that you should follow to answer me.

You are going to receive a programming problem description. The problem description begins with the Scenario Section that describes the problem, it is then followed by the Input Format Section that describes the input of the problem, then by the Output Format Section that describes the output of the problem, then by an optional Example Section that may have an example, and then by an optional Limits Section that describes the problem's memory limits, time limits, and code complexity limits if any, then by the Optimal Solution Section which contains the problem's optimal solution, and finally by an an optional Extra Code Section that contains any extra code that is required to test the problem's solution. You are required to respond with the description of edge cases that test the problem based on its input limits and constraints that are present in the Input Format and the Limits Sections. With each edge case description, generate a plain text example for that edge case.

After you generate the edge cases, please reflect on them to make sure they cover all edge cases, and double check that each plain text example actually conforms to its edge case description. If there are any mistakes, point them out.

After you finish the reflection, please take into account the feedback from the reflection, and fix any issues in the edge cases or their plain text examples.

Finally, provide me with Python3 code that generates a file called "test_cases.json" that contains native JSON output that includes all the improved edge cases in addition to another 100 completely random test cases, if possible. The input of the test cases the code will generate must follow the format defined in the Input Format Section and as can be inferred from the optimal solution. Separate the inputs of the same case with new lines or spaces as appropriate, do not use commas or other characters unless it is described in the Input Section Format as part of the input. The inputs should also be within any limits defined in the Input Format Section in the problem description. The native JSON must start with an array at the root that contains the cases. Each case has only one key: \"input\" which is a string that contains all the input. The code should not include the output of the case in the JSON. Attempt to run the code to make sure it has no errors. If you find any errors, fix them and provide the final working code.
\end{lstlisting}

\begin{lstlisting}[frame=single,breaklines=true,breakindent=0pt,breakatwhitespace=true,caption={User Prompt.\label{lst:listingc5}}]
Scenario Section:

The program is designed to take a single letter from the alphabet as input from the user and display the next letter in the alphabet sequence. The output format includes the original letter entered by the user, followed by a stylized arrow '->', and then the next letter in the alphabet. This program is intended to help users quickly find the subsequent letter of any given letter in the alphabet, with the exception that the input should not be 'z' or 'Z'.

Input Format Section:

A single letter of the alphabet (excluding 'z' or 'Z') entered by the user.

Output Format Section:

The original letter entered by the user, followed by a stylized arrow '->', and then the next letter in the alphabet.

Example Section:

Input: A
Output: A -> B

Limits Section:

Not Available

Optimal Solution Section:
  
```
#define _CRT_SECURE_NO_WARNINGS
#include <stdio.h>

int main(void)
{
  char input;
  char result;

  printf("Enter letter:\n");
  scanf("%c", &input);

  result = (char)(input + 1);

  printf("%c -> %c", input, result);

  return 0;
}
```

Extra Code Section:
```

```
\end{lstlisting}

\begin{lstlisting}[frame=single,breaklines=true,breakindent=0pt,breakatwhitespace=true,caption={LLM Response.\label{lst:listingc6}}]
### Edge Case Descriptions and Examples

1. **Edge Case: First letter of the alphabet (lowercase)**
   - Description: The input is the first letter of the alphabet in lowercase ('a'), which should return the next letter ('b').
   - Example: 
     ```
     input: a
     ```

2. **Edge Case: First letter of the alphabet (uppercase)**
   - Description: The input is the first letter of the alphabet in uppercase ('A'), which should return the next letter ('B').
   - Example:
     ```
     input: A
     ```

3. **Edge Case: Letter immediately before 'z' (lowercase)**
   - Description: The input is the letter immediately before 'z' in lowercase ('y'), which should return the next letter ('z').
   - Example:
     ```
     input: y
     ```

4. **Edge Case: Letter immediately before 'Z' (uppercase)**
   - Description: The input is the letter immediately before 'Z' in uppercase ('Y'), which should return the next letter ('Z').
   - Example:
     ```
     input: Y
     ```

5. **Edge Case: Middle of the alphabet (lowercase)**
   - Description: The input is a letter from the middle of the alphabet in lowercase ('m'), which should return the next letter ('n').
   - Example:
     ```
     input: m
     ```

6. **Edge Case: Middle of the alphabet (uppercase)**
   - Description: The input is a letter from the middle of the alphabet in uppercase ('M'), which should return the next letter ('N').
   - Example:
     ```
     input: M
     ```

### Reflection

Upon reflection, the edge cases listed above cover:
- The beginning of the alphabet in both lowercase and uppercase.
- The end of the valid input range, just before 'z' and 'Z'.
- A middle point in the alphabet to ensure the program works for letters not at the extremes.

Each plain text example accurately matches its description, adhering to the input format of a single letter of the alphabet (excluding 'z' or 'Z').

### Adjustments Based on Reflection

No adjustments are needed as the edge cases and their examples accurately reflect the constraints and requirements of the problem description.

### Python Code to Generate `test_cases.json`

```python
import json
import random
import string

def generate_random_cases(n):
  valid_letters = string.ascii_letters.replace('z', '').replace('Z', '')
  return [random.choice(valid_letters) for _ in range(n)]

def main():
  edge_cases = ['a', 'A', 'y', 'Y', 'm', 'M']
  random_cases = generate_random_cases(100)
  all_cases = edge_cases + random_cases

  test_cases = [{"input": case + "\n"} for case in all_cases]

  with open("test_cases.json", "w") as f:
    json.dump(test_cases, f, indent=4)

if __name__ == "__main__":
  main()
```

This Python script generates a file named `test_cases.json` containing the edge cases described above and an additional 100 random test cases, following the input format section's requirements.
\end{lstlisting}

\clearpage
\subsection{Function-implementation Problem Prompt\label{app3.3}}

\begin{lstlisting}[frame=single,breaklines=true,breakindent=0pt,breakatwhitespace=true,caption={System Prompt.\label{lst:listingc7}}]
Here is a step-by-step guide that you should follow to answer me.

You are going to receive a programming problem description. The problem description begins with the Scenario Section that describes the problem, it is then followed by the Input Format Section that describes the input of the problem, then by the Output Format Section that describes the output of the problem, then by an optional Example Section that may have an example, and then by an optional Limits Section that describes the problem's memory limits, time limits, and code complexity limits if any, and finally by the Optimal Solution Section which contains the problem's optimal solution. You are required to respond with the description of edge cases that test the problem based on its input limits and constraints that are present in the Input Format and the Limits Sections.

After you generate the edge case descriptions, please reflect on them to make sure they cover all edge cases. If there are any mistakes, point them out.

After you finish the reflection, please take into account the feedback from the reflection, and fix any issues in the edge cases.

Finally, provide C code that tests the problem. The code should be generated using the following template. Do not remove any code from the template, just add to it:

```
#include <stdio.h>
#include <ctype.h>
#include <string.h>
#include <stdbool.h>
#include <math.h>
#include <limits.h>
#include <time.h>
#include <stdlib.h>

#define NUM_ROWS 4
#define NUM_COLS 4

int main() {
  srand(time(NULL));

  // Add the tests here

  return 0;
}
```

1. Start by adding code to the main function. You should add tests that cover all the improved edge cases. Make sure you write each test in its own scope inside curly brackets { and }. Make sure you write all the improved edge cases fully without any placeholders. Each test case must start with:
```
printf("Test starting\n");
{
```
and must end with:
```
}
printf("Test ending\n");
```
2. Each test must call the optimal solution function. If the optimal solution function's return type is not void, print its output using printf on a separate line.
3. If the function accepts pointer parameters, the test case should also print the values of the variables that were sent to the function.
4. After the edge cases you generated, generate another 100 completely random test cases, if possible. The input of the random test cases you generate must be within any limits defined in the Input Format Section or the Limits Section in the problem description. The random tests must also be inside their own scope in curly brackets { and } and start with:
  ```
printf("Random tests starting\n");
{
```
and must end with:
```
}
printf("Random tests ending\n");
```
5. Do not forget to include any additional libraries you use in the code at the top of the file.
6. Do not attempt to add the optimal solution to the code, it will be later provided inside an included header file "solution.c".
7. Do not put any placeholders for me to fill out. The code you generate must have all the cases generated by you.
\end{lstlisting}

\begin{lstlisting}[frame=single,breaklines=true,breakindent=0pt,breakatwhitespace=true,caption={User Prompt.\label{lst:listingc8}}]
Scenario Section:

In a 2D board game, each tile on the board is represented by a unique integer. The goal is to find the position of a specific tile on the board. The board is defined by a fixed number of rows and columns. The function FindTile() is designed to locate the specified tile and update the given pointers to reflect the row and column position of the tile on the board.

Input Format Section:

The function takes four inputs: 1. A 2D array 'board' representing the game board, where each element is an integer corresponding to a tile. 2. An integer 'tile' representing the tile of interest. 3. An integer pointer 'rowPos' to store the row position of the tile. 4. An integer pointer 'colPos' to store the column position of the tile.

Output Format Section:

The function does not return any value. Instead, it updates the values pointed to by 'rowPos' and 'colPos' to reflect the row and column positions of the specified tile on the board.

Example Section:

Given a board: [[1, 2, 3, 4], [5, 6, 7, 8], [9, 10, 11, 12], [13, 14, 15, 16]] and the tile of interest is 11. After executing FindTile(), 'rowPos' should point to 2 and 'colPos' should point to 2, indicating that tile 11 is located at the third row and third column of the board (considering 0-based indexing).

Limits Section:

Not Available

Optimal Solution Section:

```
void FindTile(int board[NUM_ROWS][NUM_COLS], int tile, int *rowPos, int *colPos)
{
	// Find the position of the tile
	int i, j, row, col;

	// Locate the tile to move
	row = 0;
	col = 0;
	for (i = 0; i < NUM_ROWS; i++) {
		for (j = 0; j < NUM_COLS; j++) {
			if (board[i][j] == tile) {
				row = i;
				col = j;
			}
		}
	}
	*rowPos = row;
	*colPos = col;
}
```
\end{lstlisting}

\begin{lstlisting}[frame=single,breaklines=true,breakindent=0pt,breakatwhitespace=true,caption={LLM Response.\label{lst:listingc9}}]
Edge Case Descriptions:
Given the problem description and requirements, let's identify the edge cases based on the Input Format and any constraints provided:

1. **Tile Not Present:** The specified tile is not present on the board. This case tests how the function handles the situation when the tile cannot be found. The expectation is that the function could set `rowPos` and `colPos` to a specific value indicating the tile is not found, such as -1.

2. **Tile at Corners of the Board:** The specified tile is located at one of the corners of the board. This tests the function's ability to correctly identify tiles located at the board's edges.

3. **Tile at the Center:** The specified tile is located in the center of the board. This is a general case but ensures that the function works correctly for tiles not at the edge.

4. **Single Row or Column Board:** The board consists of a single row or column. This edge case tests the function's capability to handle non-square boards, assuming the board could be dynamically sized (even though the problem specifies a fixed size, this theoretical edge case checks flexibility).

5. **Maximum Size Board with Maximum Integer Values:** A board filled with the maximum integer values (INT_MAX) except for one tile, testing the function's ability to handle extreme integer values.

Reflecting on these edge cases, they cover a wide range of scenarios that could potentially reveal flaws in the function's logic, such as boundary issues, handling of non-existent elements, and extreme values. However, there's a mistake in suggesting a board with a single row or column since the problem specifically mentions a fixed size board defined by `NUM_ROWS` and `NUM_COLS`. Therefore, the edge case related to dynamic board sizes is not applicable here. Additionally, the edge case for a maximum size board with maximum integer values might also be irrelevant due to the fixed-size nature of the board and the lack of mention of integer limits in the problem statement.

Correcting these reflections, the improved and applicable edge cases for testing would be:

1. **Tile Not Present:** Verify function behavior when the tile is not on the board.
2. **Tile at Corners of the Board:** Ensure correct identification of tiles at the board's corners.
3. **Tile at the Center:** Test for a tile located at the center of the board.

Now, I will generate the C code to test these edge cases along with additional random test cases within the specified board limits.

```c
#include <stdio.h>
#include <ctype.h>
#include <string.h>
#include <stdbool.h>
#include <math.h>
#include <limits.h>
#include <time.h>
#include <stdlib.h>

#define NUM_ROWS 4
#define NUM_COLS 4

void printBoard(int board[NUM_ROWS][NUM_COLS]) {
    for (int i = 0; i < NUM_ROWS; ++i) {
        for (int j = 0; j < NUM_COLS; ++j) {
            printf("%d ", board[i][j]);
        }
        printf("\n");
    }
}

int main() {
    srand(10);
    int board[NUM_ROWS][NUM_COLS] = {
        {1, 2, 3, 4},
        {5, 6, 7, 8},
        {9, 10, 11, 12},
        {13, 14, 15, 16}
    };
    int rowPos, colPos;
    
    // Test for tile not present
    printf("Test starting\n");
    {
        int tile = 17; // Tile not present on the board
        FindTile(board, tile, &rowPos, &colPos);
        if (rowPos == -1 && colPos == -1) {
            printf("Tile %d not found as expected.\n", tile);
        } else {
            printf("Unexpected result for tile %d at position [%d, %d].\n", tile, rowPos, colPos);
        }
    }
    printf("Test ending\n");
    
    // Test for tile at corners of the board
    printf("Test starting\n");
    {
        int tiles[] = {1, 4, 13, 16}; // Tiles at each corner
        for (int i = 0; i < sizeof(tiles)/sizeof(tiles[0]); ++i) {
            FindTile(board, tiles[i], &rowPos, &colPos);
            printf("Tile %d found at position [%d, %d].\n", tiles[i], rowPos, colPos);
        }
    }
    printf("Test ending\n");
    
    // Test for tile at the center
    printf("Test starting\n");
    {
        int tile = 11; // Tile at the center
        FindTile(board, tile, &rowPos, &colPos);
        printf("Tile %d found at position [%d, %d].\n", tile, rowPos, colPos);
    }
    printf("Test ending\n");
    
    // Additional random tests
    printf("Random tests starting\n");
    {
        for (int i = 0; i < 100; ++i) {
            int randomTile = rand() % (NUM_ROWS * NUM_COLS) + 1; // Ensure tile is within board limits
            FindTile(board, randomTile, &rowPos, &colPos);
            printf("Random tile %d found at position [%d, %d].\n", randomTile, rowPos, colPos);
        }
    }
    printf("Random tests ending\n");
    
    return 0;
}
```

This code starts by including the necessary libraries and assumes the existence of a file named `solution.c` that contains the optimal solution. The `main` function initializes the board and defines a set of test cases, including tests for a tile not present on the board, tiles at the corners of the board, and a tile at the center of the board. It also includes 100 random tests for various tiles, ensuring that they are within the limits of the board. Each test case is wrapped in its own scope and prints the starting and ending messages as specified.
\end{lstlisting}

\clearpage

\section{Valid Solutions on Invalid Tests\label{app4}}

\setcounter{lstlisting}{0}
\setcounter{table}{0}

\subsection{Problem 2\label{app4.1}}

The reference solution in Listing \ref{lst:listingd1} sets the default row and column indices for the source to be 0, then tries to find the actual source indices by iterating over the 2D array and replacing the default values.

\begin{lstlisting}[caption={Problem 2's reference solution.\label{lst:listingd1}},,basicstyle=\fontsize{8}{10}\selectfont\ttfamily,frame=single,breaklines=true,breakindent=0pt,breakatwhitespace=true,
  postbreak=\mbox{\textcolor{red}{$\hookrightarrow$}\space},]
int Min(int a, int b) {
  if (a < b) {
    return a;
  } else {
    return b;
  }
}
int Max(int a, int b) {
  if (a > b) {
    return a;
  } else {
    return b;
  }
}
void PelletRoute(int warehouse[10][10]) {
  int i, j;
  int srow, scol;
  int erow, ecol;
  int startIndex, endIndex;
  // Locate positions
  srow = 0;
  scol = 0;
  erow = 0;
  ecol = 0;
  for (i = 0; i < 10; i++) {
    for (j = 0; j < 10; j++) {
      if (warehouse[i][j] == 1) {
        srow = i;
        scol = j;
      } else if (warehouse[i][j] == 2) {
        erow = i;
        ecol = j;
      }
    }
  }
  // Draw horizontal lines
  startIndex = Min(scol, ecol);
  endIndex = Max(scol, ecol);
  for (i = startIndex; i <= endIndex; i++) {
    if (warehouse[srow][i] == 0) {
      warehouse[srow][i] = 3;
    }
  }
  // Draw vertical lines
  startIndex = Min(srow, erow);
  endIndex = Max(srow, erow);
  for (i = startIndex; i <= endIndex; i++) {
    if (warehouse[i][ecol] == 0) {
      warehouse[i][ecol] = 3;
    }
  }
}
\end{lstlisting}

The invalid test case in Table \ref{tab:tablec1} does not have a source in it because the destination (value 2) will overwrite the source (value 1). The code will assume that the source indices are (0, 0) per the default values for the source row and source column indices set in the code, and will produce the following expected output.

\begin{table*}%
\caption{An invalid test case generated by the LLM for Problem 2 and the output of the reference solution in Listing \ref{lst:listingd1} for it.\label{tab:tablec1}}
\begin{tabular*}{\textwidth}{p{0.5\textwidth}|l}\toprule
\textbf{Test Case} & \textbf{Expected Output} \\ \midrule
\begin{tabular}[t]{@{}l}warehouse[5][5] = 1; \\ warehouse[5][5] = 2; \\ PelletRoute(warehouse);\end{tabular}
& 
\begin{tabular}[t]{@{}l}3 3 3 3 3 3 0 0 0 0 \\ 0 0 0 0 0 3 0 0 0 0 \\ 0 0 0 0 0 3 0 0 0 0 \\ 0 0 0 0 0 3 0 0 0 0 \\ 0 0 0 0 0 3 0 0 0 0 \\ 0 0 0 0 0 2 0 0 0 0 \\ 0 0 0 0 0 0 0 0 0 0 \\ 0 0 0 0 0 0 0 0 0 0 \\ 0 0 0 0 0 0 0 0 0 0 \\ 0 0 0 0 0 0 0 0 0 0 \end{tabular} \\
\bottomrule
\end{tabular*}
\end{table*}

The solution in Listing \ref{lst:listingd2} is a valid student solution, it sets the default source row and column indices to 0 then tries to find the actual indices, just like the reference solution did. The invalid test case in Table \ref{tab:tablec1} will cause this code to assume that the source is in indices (0, 0), and will therefore produce the same output as the reference solution as shown in Table \ref{tab:tablec2}, and that will make it pass the LLM Test Suite.

\begin{lstlisting}[caption={A valid student solution for problem 2.\label{lst:listingd2}},basicstyle=\fontsize{8}{10}\selectfont\ttfamily,frame=single,breaklines=true,breakindent=0pt,breakatwhitespace=true,
  postbreak=\mbox{\textcolor{red}{$\hookrightarrow$}\space},]
#include <stdio.h>

void PelletRoute(int warehouse[10][10])
{
	int i, j, p1row, p1col, p2row, p2col;
	p1row = 0;
	p1col = 0;
	p2row = 0;
	p2col = 0;

	for (i = 0; i < 10; i++) {
		for (j = 0; j < 10; j++) {
			if (warehouse[i][j] == 1) {
				p1row = i;
				p1col = j;
			}
			if (warehouse[i][j] == 2) {
				p2row = i;
				p2col = j;
			}
		}
	}

	int rowDiff, colDiff, across, down;

	rowDiff = p2row - p1row;
	colDiff = p2col - p1col;

	if (colDiff > 0) {
		for (across = 0; across < colDiff; across++) {
			if (warehouse[p1row][p1col + across] != 1) {
				warehouse[p1row][p1col + across] = 3;
			}
		}
	}

	if (colDiff < 0) {
		colDiff = -colDiff;
		for (across = 0; across < colDiff; across++) {
			if (warehouse[p1row][p1col - across] != 1) {
				warehouse[p1row][p1col - across] = 3;
			}
		}
	}


	if (rowDiff > 0) {
		for (down = 0; down < rowDiff; down++) {
			if (warehouse[p1row + down][p2col] != 1) {
				warehouse[p1row + down][p2col] = 3;
			}
		}
	}

	if (rowDiff < 0) {
		rowDiff = -rowDiff;
		for (down = 0; down < rowDiff; down++) {
			if (warehouse[p1row - down][p2col] != 1) {
				warehouse[p1row - down][p2col] = 3;
			}
		}
	}

}
\end{lstlisting}

\begin{table*}[h!]%
\caption{The outputs of the solution in Listing \ref{lst:listingd2} on the test case in Table \ref{tab:tablec1}.\label{tab:tablec2}}
\begin{tabular*}{\textwidth}{p{0.1\textwidth}|l}\toprule
\textbf{Output}
& 
\begin{tabular}{@{}l}3 3 3 3 3 3 0 0 0 0 \\ 0 0 0 0 0 3 0 0 0 0 \\ 0 0 0 0 0 3 0 0 0 0 \\ 0 0 0 0 0 3 0 0 0 0 \\ 0 0 0 0 0 3 0 0 0 0 \\ 0 0 0 0 0 2 0 0 0 0 \\ 0 0 0 0 0 0 0 0 0 0 \\ 0 0 0 0 0 0 0 0 0 0 \\ 0 0 0 0 0 0 0 0 0 0 \\ 0 0 0 0 0 0 0 0 0 0 
\end{tabular} \\
\bottomrule
\end{tabular*}
\end{table*}

Though, the valid solution in Listing \ref{lst:listingd3} sets the default row and column indices to (-1, -1). The invalid test case in Table \ref{tab:tablec1} does not have a source in it so the code will keep the values (-1, -1) as the indices of the source. As soon as the code uses the subscript operator ``warehouse[pRow][pCol]'' the first time, the code will throw a runtime error because pRow and pCol will both be -1. This will cause the solution to fail the LLM Test Suite even though it is a valid solution.

\begin{lstlisting}[caption={Another valid student solution for problem 2.\label{lst:listingd3}},basicstyle=\fontsize{8}{10}\selectfont\ttfamily,frame=single,breaklines=true,breakindent=0pt,breakatwhitespace=true,
  postbreak=\mbox{\textcolor{red}{$\hookrightarrow$}\space},]
void PelletRoute(int warehouse[10][10]) {
    int pRow = -1, pCol = -1, dRow = -1, dCol = -1;

    for (int i = 0; i < 10; ++i) {
        for (int j = 0; j < 10; ++j) {
            if (warehouse[i][j] == 1) {
                pRow = i;
                pCol = j;
            }
            else if (warehouse[i][j] == 2) {
                dRow = i;
                dCol = j;
            }
        }
    }

    while (pCol != dCol) {
        if (pCol < dCol) {
            ++pCol;
        }
        else {    
            --pCol;
        }
        if (warehouse[pRow][pCol] != 2) {
            warehouse[pRow][pCol] = 3;
        }
    }

    while (pRow != dRow) {
        if (pRow < dRow) {
            ++pRow;
        }
        else {
            --pRow;
        }
        if (warehouse[pRow][pCol] != 2) {
            warehouse[pRow][pCol] = 3;
        }
    }
}
\end{lstlisting}

\begin{table*}[h!]%
\caption{The outputs of the solution in Listing \ref{lst:listingd3} on the test case in Table \ref{tab:tablec1}.\label{tab:tablec3}}
\begin{tabular*}{\textwidth}{p{0.1\textwidth}|l}\toprule
\textbf{Output}
& 
Runtime Error \\
\bottomrule
\end{tabular*}
\end{table*}

\subsection{Problem 16\label{app4.2}}

The reference solution in Listing \ref{lst:listingd4} goes over all the letters in a string and replaces the letters between left and right with 'X'.

\begin{lstlisting}[caption={Problem 16's reference solution.\label{lst:listingd4}},basicstyle=\fontsize{8}{10}\selectfont\ttfamily,frame=single,breaklines=true,breakindent=0pt,breakatwhitespace=true,
  postbreak=\mbox{\textcolor{red}{$\hookrightarrow$}\space},]
void Censor(char *phrase, int left, int right)
{
	int i = 0;
	while (phrase[i] != '\0') {
		if (i >= left && i <= right) {
			phrase[i] = 'X';
		}
		i++;
	}
}
\end{lstlisting}

The test case in Table \ref{tab:tablec4} is invalid according to the problem statement because the first index (7) is not less than or equal to the second index (3), and the problem statement does not specify how to handle bad range indices. The reference solution in Listing \ref{lst:listingd4} will not enter the if-statement inside the while-loop and therefore the string will not be modified. The printf statement in the test case will print the string as is.

\begin{table*}[h!]%
\caption{An invalid test case generated by the LLM for Problem 16 and the output of the reference solution in Listing \ref{lst:listingd4} for it.\label{tab:tablec4}}
\begin{tabular*}{\textwidth}{p{0.5\textwidth}|l}\toprule
\textbf{Test Case} & \textbf{Expected Output} \\ \midrule
\begin{tabular}[t]{@{}l}char phrase[] = "Hello World"; \\ Censor(phrase, 7, 3); \\ printf("\%s\textbackslash n", phrase);
\end{tabular}
& 
Hello World \\
\bottomrule
\end{tabular*}
\end{table*}

The solution in Listing \ref{lst:listingd5} is a valid student solution, it iterates the letters of the string with a for-loop and changes the letters that are in the range [left, right] to 'X'. Similar to the reference solution in Listing \ref{lst:listingd4}, the code will not enter the if-statement and will not modify the string. The printf statement in the test case will print the string as is, and so the solution will pass the LLM Test Suite because it matches the expected output as shown in Table \ref{tab:tablec5}.
\vspace{12pt}
\begin{lstlisting}[caption={A valid student solution for problem 16.\label{lst:listingd5}},basicstyle=\fontsize{8}{10}\selectfont\ttfamily,frame=single,breaklines=true,breakindent=0pt,breakatwhitespace=true,
  postbreak=\mbox{\textcolor{red}{$\hookrightarrow$}\space},]
#define _CRT_SECURE_NO_WARNINGS
#include <stdio.h>
#include <stdlib.h>
#include <string.h>

void Censor(char* phrase, int left, int right)
{
	int i;

	for (i = 0; i < strlen(phrase); i++) {
		if (i >= left && i <= right) {
			phrase[i] = 88;
		}
	}

	return;
}
\end{lstlisting}

\begin{table*}[h!]%
\caption{The outputs of the solution in Listing \ref{lst:listingd5} on the test case in Table \ref{tab:tablec4}.\label{tab:tablec5}}
\begin{tabular*}{\textwidth}{p{0.1\textwidth}|l}\toprule
\textbf{Output}
& 
Hello World \\
\bottomrule
\end{tabular*}
\end{table*}

On the other hand, the student that wrote the valid solution in Listing \ref{lst:listingd6} decided to validate the range, and print a message in case it is invalid. The extra output will cause the solution to fail the LLM Test Suite even though the solution itself is valid as shown in Table \ref{tab:tablec6}.

\begin{lstlisting}[caption={Another valid student solution for problem 16.\label{lst:listingd6}},basicstyle=\fontsize{8}{10}\selectfont\ttfamily,frame=single,breaklines=true,breakindent=0pt,breakatwhitespace=true,
  postbreak=\mbox{\textcolor{red}{$\hookrightarrow$}\space},]
void Censor(char str[], int left, int right) {
    int length = strlen(str);
    if (left < 0 || right >= length || left > right) {
        printf("Invalid indices\n");
        return;
    }
    for (int i = left; i <= right; i++) {
        str[i] = 'X';
    }
}
\end{lstlisting}

\begin{table*}[h!]%
\caption{The outputs of the solution in Listing \ref{lst:listingd6} on the test case in Table \ref{tab:tablec4}.\label{tab:tablec6}}
\begin{tabular*}{\textwidth}{p{0.1\textwidth}|l}\toprule
\textbf{Output}
& 
\begin{tabular}{@{}l}Invalid indices \\ Hello World
\end{tabular} \\
\bottomrule
\end{tabular*}
\end{table*}

\end{document}